\documentstyle[12pt,psfig]{article}
\def\dfrac#1#2{{\displaystyle\frac{#1}{#2}}}
\newcommand{\lsim}{\raisebox{0.3mm}{\em $\, <$}
\hspace{-3.3mm} \raisebox{-1.8mm}{\em $\sim \,$}}
\newcommand{\gsim}{\raisebox{0.3mm}{\em $\, >$}
\hspace{-3.3mm} \raisebox{-1.8mm}{\em $\sim \,$}}
\begin{document}
\makeatletter
%
%
%
%
%
\newtoks\@stequation

\def\subequations{\refstepcounter{equation}%
  \edef\@savedequation{\the\c@equation}%
  \@stequation=\expandafter{\theequation}
  \edef\@savedtheequation{\the\@stequation}
  \edef\oldtheequation{\theequation}%
  \setcounter{equation}{0}%
  \def\theequation{\oldtheequation\alph{equation}}}

\def\endsubequations{%
  \ifnum\c@equation < 2 \@warning{Only \the\c@equation\space  
subequation
    used in equation \@savedequation}\fi
  \setcounter{equation}{\@savedequation}%
  \@stequation=\expandafter{\@savedtheequation}%
  \edef\theequation{\the\@stequation}%
  \global\@ignoretrue}
\makeatother
\vglue -1.5cm
\rightline{FTUV/96-72, IFIC/96-81}
\rightline{October 1996}
\rightline{astro-ph/9610209}
\vglue 0.5cm
\begin{center}
\Large\bf Resonant Neutrino Spin-Flavor Precession and\\
Supernova Nucleosynthesis and Dynamics\\
\end{center}
\vglue 0.8cm
\centerline{\bf
H.  Nunokawa$^{1}$, Y.-Z. Qian$^{2}$, and G. M. Fuller$^{3}$}
\vglue 0.5cm
\centerline{\it Institute for Nuclear Theory, University of  
Washington,}
\centerline{\it Box 351550, Seattle, WA 98195}
\vglue 0.5cm
\centerline{$^1$\it Present Address: Instituto de F\'{\i}sica 
Corpuscular --- C.S.I.C.}
\centerline{\it Departament de F\'{\i}sica Te\`orica, Universitat 
de Val\`encia}
\centerline{\it 46100 Burjassot, Val\`encia, SPAIN  }
\vglue 0.5cm
\centerline{$^2$\it Present Address: Physics Department, 161-33,  
California Institute of Technology,}
\centerline{\it Pasadena, CA 91125}
\vglue 0.5cm
\centerline{$^3$\it Permanent Address: Department of Physics,  
University
of California, San Diego,}
\centerline{\it La Jolla, CA 92093-0319}
\vglue 1.3cm
\centerline{ABSTRACT}
\vglue 0.5cm
We discuss the effects of resonant spin-flavor
precession (RSFP) of Majorana neutrinos
on heavy element nucleosynthesis in neutrino-heated
supernova ejecta and the  
dynamics of supernovae.
In assessing the effects of RSFP,  
we explicitly include matter-enhanced (MSW) resonant neutrino flavor  
conversion effects where appropriate. We point out that for plausible
ranges of neutrino magnetic moments and proto-neutron star
magnetic fields, spin-flavor conversion of
$\nu_\tau$ (or $\nu_\mu$) with a cosmologically significant
mass (1--100 eV) into a light $\bar \nu_e$ could lead to an enhanced
neutron excess in neutrino-heated supernova ejecta.
This could be beneficial for models of $r$-process nucleosynthesis
associated with late-time neutrino-heated ejecta from supernovae.  
Similar spin-flavor conversion of neutrinos
at earlier epochs could lead to an increased shock
reheating rate and, concomitantly, a larger supernova explosion
energy. We show, however, that such increased
neutrino heating likely will be
accompanied by an enhanced neutron excess which could exacerbate  
the problem of the
overproduction of the neutron number $N = 50$ nuclei in the
supernova ejecta from this  
stage. In all of these scenarios, the average $\bar\nu_e$ 
energy will be increased over those predicted by supernova models  
with no neutrino mixings. This may allow the SN1987A data to  
constrain RSFP-based schemes. 

\newpage
\noindent
\renewcommand{\thesection}{\Roman{section}.}
\centerline{\bf I. Introduction}
\vglue 0.5cm
In this paper we examine the effects of Resonant Spin-Flavor  
Precession (RSFP) of Majorana neutrinos on 
heavy element nucleosynthesis in neutrino-heated supernova
ejecta and the dynamics of
supernovae. Massive neutrinos 
that possess a transition magnetic  
moment could experience a resonant conversion of spin and flavor in  
the presence of magnetic fields and matter \cite{LM, Ak}. This RSFP  
effect is similar to the well-known Mikheyev-Smirnov-Wolfenstein  
(MSW) mechanism \cite{MSW}. In fact, in the environment above the  
neutrinosphere in post-core-bounce supernovae, both the MSW and RSFP  
processes may operate. Reference \cite{ALPS} discusses the effect of  
RSFP-induced neutrino flavor conversion on supernova shock reheating.  
Here we extend their discussion by considering the effects of RSFP on  
nucleosynthesis in late-time neutrino-heated 
supernova ejecta and generally  
including the effects of {\it both} RSFP {\it and} MSW conversion.  

Supernovae have long been considered as a promising site for
heavy element nucleosynthesis \cite{Burbidge}.
About one half of the elements with mass number $A>70$
in nature are believed to be made by the
rapid neutron capture process, or $r$-process
for short. In the $r$-process, neutron captures occur
much faster than typical $\beta$-decays.
Recent calculations suggest that the
high-temperature, high-entropy region which
would be formed 
above the proto-neutron star
several seconds after the bounce of the core
in a type-II supernova is
a promising site for $r$-process nucleosynthesis
\cite{Woosley1,Woosley2,Woosley3}.

This site is sometimes referred to as the ``hot bubble,'' since
material there is heated by
absorption of neutrinos and antineutrinos which
are emitted from the neutrinosphere near the surface of the
hot proto-neutron star.
Close to the neutrinosphere the temperature is high enough
that all strong and electromagnetic nuclear interactions
are in equilibrium (nuclear statistical equilibrium,
or NSE).
As the material above the neutrinosphere expands
due to neutrino heating, its temperature and density
decrease. When the temperature drops below about 0.5 MeV,
the material outflow rate (or expansion  
rate) becomes
faster than the rates for nuclear reactions.
At this point, the material freezes out of NSE.  
As the material further expands above this nuclear freeze-out point,
rapid neutron captures onto the existing seed nuclei
may occur after an alpha-rich freeze-out 
of the charged-particle reactions. 

In order for the $r$-process to occur,
the material certainly has to be neutron rich
at the nuclear freeze-out position.
The neutron-to-proton ratio 
above the neutrinosphere is determined by
the following reactions,
\begin{subequations}
\begin{eqnarray}
\nu_e + n       &\rightarrow \ p + e^{-},  \label{eqn:beta1a}\\
{\bar\nu_e} + p &\rightarrow \ n + e^{+}.
\label{eqn:beta1b}
\end{eqnarray}
\end{subequations}
Because material in the surface layers of the proto-neutron star
consists mostly of neutrons, 
$\nu_e$ have a larger opacity than $\bar\nu_e$ [cf. Eqs. 
(\ref{eqn:beta1a}) and (\ref{eqn:beta1b})].   
Consequently,
$\bar \nu_e$ decouple  
deeper inside the proto-neutron star where it is hotter, and
correspondingly have a higher average energy than $\nu_e$.
In turn, this average energy hierarchy favors  
the rate of the process in Eq. (\ref{eqn:beta1b})
over that in Eq. (\ref{eqn:beta1a}). 
These arguments suggest that neutron-rich conditions which  
are conducive to $r$-process nucleosynthesis
will obtain at the nuclear freeze-out point and beyond, 
in the region where the  
$r$-process may occur.

Reference \cite{QFMMW} discusses these weak  
reaction issues and the
connection between the flavor mixing of
neutrinos with cosmologically significant masses
and the conditions necessary for heavy element nucleosynthesis
in supernovae. They show that resonant flavor conversion of
a $\nu_\tau$ or $\nu_\mu$ with a mass of 1--100 eV into a light
$\nu_e$ may preclude heavy element nucleosynthesis in the hot bubble  
unless
the vacuum mixing angle satisfies $\sin^2 2\theta \lsim 10^{-5}$.
Because $\nu_\mu$ and $\nu_\tau$ and their antineutrinos
lack the charged-current reactions similar to those in Eqs.
(\ref{eqn:beta1a}) and (\ref{eqn:beta1b}), they decouple deepest
inside the proto-neutron star and have the highest average energy.
Consequently, a significant amount of $\nu_{\tau(\mu)}
\leftrightarrow \nu_e$ transformation results in 
$\nu_e$ with an average energy higher than that of $\bar\nu_e$. 
This enhances the rate of the process in Eq.  
(\ref{eqn:beta1a}) and
drives the neutrino-heated supernova ejecta proton-rich. 

If the mass of $\nu_\tau$ (or $\nu_\mu $) is between about 1  
and 100 eV,
then matter-enhanced resonant MSW flavor conversion could occur
in the region between the neutrinosphere and the 
radius where the weak reactions in Eqs. (\ref{eqn:beta1a})
and (\ref{eqn:beta1b}) 
freeze out (very near the nuclear freeze-out position in most  
supernova models). If average energy $\nu_\mu$ or $\nu_\tau$ are  
converted at resonance with greater than about (25--30)\%  
efficiency, then the hot bubble will be driven proton-rich and  
$r$-process nucleosynthesis at this site will be impossible.

With no such flavor conversion, the neutron excess required for  
$r$-process nucleosynthesis may or may not be obtained currently in  
models of neutrino-heated supernova ejecta. Although the conditions  
of entropy, electron fraction [a measure of the
neutron-to-proton ratio and the neutron excess, cf. Eq. (\ref{Ye})], 
and expansion rate as computed in some  
hydrodynamic models \cite{Woosley3} provide the requisite  
neutron-to-seed ratio for the $r$-process to occur, these  
models have left out key physics input that can wreck the $r$-process  
\cite{Meyer95}. Besides, simple wind model arguments suggest that the  
neutron-to-seed ratio obtained in these hydrodynamic models  
is unrealistically large on account of their high entropies at late  
times. Wind models suggest an entropy roughly half of that obtained  
in some hydrodynamic models, and this would imply a neutron-to-seed  
ratio too low to allow the production of the heaviest  
$r$-process nuclides \cite{QW96,MLB96}. 

Any effect that could {\it lower} the electron fraction $Y_e$ (i.e.,  
raise the neutron-to-seed ratio) in these models would be  
most welcome. In fact, it is even conceivable that $r$-process  
nucleosynthesis in late-time neutrino-heated supernova ejecta will be  
impossible unless there is some new physics which has the effect of  
raising the neutron excess \cite{FQW96,Qian96}.
However, any effect that significantly lowers $Y_e$ at late times  
must not also do so at the early times characteristic of shock  
reheating, lest the neutron number $N=50$ nuclei be overproduced  
\cite{Woosley3,FM95,Hoffman96}. This shock reheating  
epoch occurs roughly $\sim 0.1$--0.6 s after core bounce, as opposed  
to the epoch where the $r$-process might take place in  
neutrino-heated ejecta at $\sim 3$--20 s after core  
bounce. 

In what follows we show that $\nu_\tau$ (or $\nu_\mu$) with
vacuum masses in the range 1--100 eV
could convert in principle into $\bar \nu_e$ by
RSFP above the neutrinosphere. In turn, 
this could lead to the {\it enhancement} of the neutron excess in the  
``hot bubble.''
In order to have the conversion $\nu_{\tau(\mu)}
\leftrightarrow \bar \nu_e$, neutrinos must be of Majorana type and
have a finite transition magnetic moment. 
The operation of RSFP also
requires a (probably large) magnetic field around the 
proto-neutron star. If RSFP did occur, $\bar \nu_e$ would become
much more energetic than $\nu_e$ and this would enlarge the  
neutron-to-proton ratio as outlined above.  This implies that RSFP  
could be beneficial to the production of the heavier $r$-process  
elements in supernovae, in contrast to
the case of resonant MSW flavor conversion alone as discussed in Ref.  
\cite{QFMMW}.
For significant RSFP effects to occur, the required product  
of the neutrino transition magnetic moment
$\mu_\nu$ and magnetic field $B$ near the neutron star will be shown  
to be of order 1 in units of Bohr magneton times gauss
($\mu_B\cdot\mbox{G}$).

Before entering into a detailed discussion
of RSFP in supernovae, let us describe the present upper limits on the
neutrino magnetic moment from
laboratory experiments and astrophysical arguments.
The present upper bound on the neutrino magnetic moment from
${\bar \nu_e}e$ scattering experiments is \cite{Derbin}
\begin{equation}
\mu_{\nu}  < 1.9\times10^{-10} \mu_B.
 \label{eqn:upperbound}
\end{equation}
This bound applies to the direct or transition magnetic moment of
Dirac neutrinos, as well as the transition magnetic moment of  
Majorana
neutrinos.

A stronger limit for the neutrino magnetic moment can be derived
from well-known arguments against excessive cooling of red giant stars.
A finite neutrino magnetic moment would enhance the
plasmon decay into $\nu \bar \nu$ pairs  in the stellar interior,
resulting in excessive cooling.
The most severe constraint is derived by estimating the critical
mass for a helium flash in red giant stars. This 
gives the bound \cite{Raffelt}
\begin{equation}
\mu_\nu < 3\times10^{-12} \mu_B
\end{equation}
for the transition magnetic moment of Majorana neutrinos.

A magnetic moment of
order $(10^{-12}$--$10^{-10})\mu_B$ is
very large in the context of the small neutrino masses we might  
consider here.
In general, this is because the diagram that generates a 
magnetic moment of order $10^{-12}\mu_B$ or larger with the 
photon line removed also induces a large neutrino mass.
There are many (successful but not compelling) 
attempts at constructing a mechanism to
induce a large neutrino magnetic moment, while
keeping the masses of neutrinos small \cite{Magneticmomentmodels}.
Since these issues are very speculative, we will assume here that a  
neutrino transition magnetic moment
of order $10^{-12}\mu_B$ is plausible.
However, we will see that the crucial quantity 
governing the effects of RSFP 
in supernovae is the product of the neutrino 
transition magnetic moment
and the magnetic field. 

\vglue 1cm
\centerline{\bf II. Resonant Neutrino Spin-Flavor Precession in Supernovae}
\vglue 0.5cm
In this section we discuss some general features of
RSFP in supernovae. We also examine the case where  
{\it both}
matter-enhanced MSW flavor conversion {\it and} RSFP occur in the  
region above the hot proto-neutron star.
Both Dirac and Majorana neutrinos can have RSFP, so
long as the transition magnetic moment exists. The
mechanism of RSFP is essentially the same for both cases.
However, the implications of RSFP for supernova heavy element  
nucleosynthesis 
and/or explosion dynamics will be very different for the  
two cases.
This is because RSFP for Majorana neutrinos occurs between
two active neutrinos, whereas RSFP for Dirac neutrinos occurs  
between
active and sterile neutrino states. In this paper we shall consider  
only the Majorana neutrino case.

The Lagrangian which describes the magnetic moment-mediated  
interaction
between Majorana neutrinos and the electromagnetic field, $F_{\alpha  
\beta}$,
is given by
\begin{equation}
{\cal L}_{\rm int}={1\over 2}(\mu)_{ab}
({\bar\nu_L})^C_a\sigma_{\alpha\beta}
(\nu_L)_b F^{\alpha\beta} +{\rm h.c.},
\label{eqn:Lagrangian}
\end{equation}
where $(\mu)_{ab}$ is the magnetic moment matrix with
$a,\ b=e,\ \mu,\ \tau$ or $ 1,\ 2,\ 3$ for
the flavor or mass eigenstate bases, 
$\sigma_{\alpha\beta}=(i/2)[\gamma_\alpha,\gamma_\beta]$
with $\gamma_\alpha$ the Dirac matrices, and $C$ denotes the
operation of charge conjugation. From the requirement for $CPT$  
invariance, the 
diagonal elements of the magnetic moment $\mu_{aa}$ vanish, 
and consequently a transition magnetic moment
is the only possibility for Majorana neutrinos.
In the case of a finite neutrino transition magnetic moment, 
the presence of magnetic fields can facilitate the
transformation 
$\nu_{a L} \rightarrow (\nu_{b L})^C$
($a \ne b$) or vice versa.
Since $(\nu_{b L})^C$ is generally termed $\bar \nu_{b}$
(antineutrino state for $\nu_{b}$)
and is right handed, we can describe the
$\nu_a \leftrightarrow {\bar \nu_{b}}\ (a \ne b)$
transformation
as a \lq\lq spin-flavor\rq\rq\  precession (or conversion).
Except for the interaction in Eq. (\ref{eqn:Lagrangian}), 
we will assume here that neutrinos 
possess only standard electroweak interactions with matter.

Among the several conceivable channels of spin-flavor precession,
$\nu_{\tau(\mu)} \leftrightarrow {\bar\nu_e}$
is the example we will consider in what follows. Motivated by the  
average neutrino energy hierarchy discussed in the last section, we  
anticipate that this channel of RSFP may give the most significant  
effect on the electron fraction. 
We will show that the spin-flavor conversion of
$\nu_\tau$ (or $\nu_\mu$) with masses in the range 1--100 eV into a  
light $\bar \nu_e$
can result in important effects on the parameters which determine
heavy element nucleosynthesis and/or shock-reheating in
the region above the neutrinosphere in supernovae.
Hereafter, we will assume in this paper that $\nu_\tau$ is
the heavy neutrino
with a mass in the range 1--100 eV, and we will  
consider the
two generation system of
electron and tau neutrinos only. Obviously, our computed effects in  
the supernova would be identical if instead we were to choose 
${\nu}_{\mu}$ as the heavy neutrino. This follows since we expect the  
energy spectra of $\nu_\mu$ and $\nu_\tau$ and their antiparticles  
to be nearly identical in our region of interest in supernovae.  
Working only with two neutrino generations is justified, so long as  
we assume
that there exists a reasonable hierarchy of the three neutrino masses  
in which no degeneracy occurs. 

The Majorana neutrino evolution equation for two neutrino  
generations,
including a vacuum mixing angle, a transition magnetic moment, and magnetic  
fields, is given by \cite{LM}
\begin{equation}
{i{d \over dr}\left[\matrix{
\nu_e \cr\ \nu_\tau \cr {\bar \nu_e}
\cr{\bar \nu_\tau}\cr}\right]=H
\left[\matrix{
\nu_e \cr\ \nu_\tau \cr {\bar \nu_e}
\cr{\bar \nu_\tau}\cr}\right] },
\label{eqn:evolution1}
\end{equation}
\begin{equation}
H=
 \left[\matrix{
& a_{\nu_e}+{\Delta} \sin^2\theta
& {1\over 2} \Delta  \sin2\theta
&\kern-0.2in 0
&\kern-0.2in\mu_\nu B_\perp \cr
&\kern-0.2in{1\over 2}\Delta \sin2\theta
& a_{\nu_\tau}+\Delta \cos^2\theta \
&\kern-0.2in -\mu_\nu B_\perp
&\kern-0.2in 0 \cr
&\kern-0.2in 0
&\kern-0.2in -\mu_\nu B_\perp
&\kern-0.2in -a_{\nu_e}+\Delta  \sin^2\theta
&{1\over 2} \Delta  \sin2\theta \cr
&\kern-0.2in\mu_\nu B_\perp
&\kern-0.2in 0
&\kern-0.2in{1\over 2} \Delta  \sin2\theta
&-a_{\nu_\tau}+\Delta \cos^2\theta \cr}
\right],
\label{eqn:Hamiltonian}
\end{equation}
\vglue 0.1cm
\noindent
where $\theta$ is the vacuum mixing angle, and
$\mu_\nu$ is the transition magnetic moment between
$\nu_e$ and ${\bar \nu_\tau}$.
Here $B_\perp$ is the transverse component of the
magnetic field along  
the neutrino
trajectory.
In the usual fashion we define  $\Delta \equiv \delta m^2 /2E_\nu$, where
$\delta m^2 \equiv m_2^2-m_1^2>0$ is the difference of the squared
vacuum mass eigenvalues
of the two neutrino mass eigenstates $\nu_2\sim \nu_\tau$ and  
$\nu_1\sim\nu_e $,
and $E_\nu$ is the neutrino energy.
We assume that the vacuum mixing angle $\theta$ is very small, so
that in vacuum with no magnetic fields the mass eigenstates are  
approximately coincident with the flavor
eigenstates. The effective matter potentials for
$\nu_e$ and $\nu_\tau$ are given by
\begin{eqnarray}
a_{\nu_e}    = &\sqrt{2} G_F\ (n_e-\frac{1}{2}n_n),\\
a_{\nu_\tau}  = & \sqrt{2} G_F\ (-\frac{1}{2}n_n),
\end{eqnarray}
where $G_F$ is the Fermi constant, $n_e$ and $n_n$ are the
net number densities of electrons and neutrons, respectively.
These expressions are for a neutral unpolarized medium and
we neglect the contribution from neutrino-neutrino scattering
because its effect would be small
under the conditions we consider here 
(see Ref. \cite{QF} for detailed studies of neutrino-neutrino 
scattering effect on MSW neutrino flavor transformation).
In Eq. (\ref{eqn:Hamiltonian}), $n_e$, $n_n$ and $B_\perp$ are all
understood to be position dependent in the region above the 
neutrinosphere which we consider here.

By equating each two of the diagonal elements in the Hamiltonian
matrix in Eq. (\ref{eqn:Hamiltonian}), we find that there are
two kinds of resonances. As expected, these correspond to MSW 
conversion and  
RSFP.
The MSW resonance occurs when
\begin{equation}
\sqrt{2}G_F\ n_e = \Delta \cos 2\theta.
\label{eqn:MSWcondition}
\end{equation}
The RSFP resonance occurs when
\begin{equation}
\sqrt{2}G_F(n_e-n_n) = \pm \Delta \cos 2\theta,
\label{eqn:RSFPcondition}
\end{equation}
where the plus sign is for the $\nu_e$-${\bar \nu_\tau}$
resonance
and the minus sign is for the ${\bar \nu_e}$-${\nu_\tau}$
resonance.
These two RSFP resonances
cannot occur at the same time.
As can be readily seen from Eq. (\ref{eqn:RSFPcondition}),
the $\nu_e$-${\bar \nu_\tau}$ resonance occurs if the sign
of $n_e-n_n$ is positive, whereas the 
${\bar \nu_e}$-${\nu_\tau}$ resonance
occurs if the sign of $n_e-n_n$ is negative at the resonance  
position.

The resonance region of most interest here lies above the  
neutrinosphere,
but, at an early epoch ($t_{\rm PB}\approx 0.1$--0.6 s), within the  
radius where the shock has stalled ($r\approx$ 500 km)
and/or, at a
later epoch ($t_{\rm PB}\approx 3$--20 s),
within the radius where the weak and/or nuclear reactions freeze out  
($r\approx$ 40 km).
Here, $t_{\rm PB}$ indicates the time {\it post core bounce}.
In the standard supernova models, $n_e-n_n$ takes negative
values in the regions of interest.
Hence, the relevant RSFP we will consider is
${\bar \nu_e} \leftrightarrow {\nu_\tau}$.
For a neutral medium ($n_e=n_p$), 
the neutron excess can be characterized by the
electron fraction $Y_e$, the net number of
electrons per baryon,
\begin{equation}
Y_e \equiv \frac{n_e}{n_e+n_n}.
  \label{Ye}
\end{equation}
If $Y_e<0.5$ then $n_n>n_e$. From numerical supernova models,
the typical values of $Y_e$ are predicted to be
about 0.4--0.45 in the region of interest above the neutrinosphere  
\cite{Wilson}.

In Fig. 1 we have plotted schematically as functions of matter  
density $\rho$ the neutrino energy levels
(effective mass-squared differences)
in neutron-rich environments ( $n_n> n_p$)
for two generations of Majorana neutrinos. To draw the curves in this  
figure
we have assumed that $1/3<Y_e < 1/2$. This condition on the electron  
fraction should be valid so long as
we confine our considerations to the region well above the  
neutrinosphere.
The number density of electrons and neutrons are
related to  matter density $\rho$ in the following manner:
\begin{eqnarray}
n_e =& \rho\  Y_e N_A, \\
n_n =& \rho\  (1-Y_e) N_A,
\end{eqnarray}
where $N_A$ is Avogadro's number.
The resonance density for RSFP is given by
\begin{equation}
\rho_{\rm res}^{\rm RSFP} \approx  6.6 \times 10^{7}
\left[\dfrac{\delta m^2 }{{  100\ \mbox{eV}}^2}\right]
 \Biggl[ \dfrac{10\  \mbox{MeV}}{E_\nu}  \Biggr]
\frac{\cos2\theta}{1-2Y_e}
\ \ {\mbox{g  \ cm}}^{-3},
\end{equation}
whereas the MSW resonance density is
\begin{equation}
\rho_{\rm res}^{\rm MSW} = \frac{1-2Y_e}{Y_e}\rho_{\rm res}^{\rm RSFP}.
\label{eqn:resodensity}
\end{equation}
As one can see from Eq. (\ref{eqn:resodensity}),
the resonance density for RSFP is larger
than that for MSW conversion as long as $1/3 < Y_e < 1/2$.
This implies that the RSFP resonance takes place {\it before} the MSW
resonance as neutrinos propagate from the neutrinosphere
to the outer regions of the supernova (see Fig. 1).
Since the matter density at the neutrinosphere is $\gsim  
10^{12}$ g cm$^{-3}$, neutrinos with typical energies and possessing  
cosmologically significant masses (1--100 eV) will propagate through  
one or more resonances.

In order to illustrate the mechanism of RSFP,
let us first work with the system
of ${\bar \nu_e}$ and ${\nu_\tau}$ alone,
and ignore for the time being MSW flavor conversion.
However, it should be noted that the following discussion of the  
effects of RSFP will be valid even if the MSW resonance were to  
coexist with
RSFP, so long as the two resonances are well separated
[see Eq. (\ref{eqn:nonoverlap}) for the
non-overlapping condition for the two resonances].
In the limit where the vacuum mixing angle is vanishingly  
small, $\theta \rightarrow 0 $,
Eq. (\ref{eqn:evolution1}) can be reduced to the following expression,
\begin{equation}
i{d \over dr}\left[\matrix{{\bar \nu_e} \cr\ {\nu_\tau}} \right]=
\left[\matrix{\ \ -a_{\nu_e} \hskip 1.0cm  -\mu_\nu B \ \ \cr
-\mu_\nu B  \hskip 0.7cm a_{\nu_\tau}+\Delta }
 \right]
\left[\matrix{{\bar \nu_e} \cr\ {\nu_\tau}} \right].
\end{equation}
In this equation and hereafter we will simply write $B$ where we have  
written $B_\perp$ before. However, it should be understood always  
that only
the transverse component of the magnetic field
along neutrino trajectories is relevant for neutrino spin-flavor  
precession.
The resonance condition for RSFP in this case is given by
\begin{equation}
\sqrt{2}G_F\ (n_n-n_e) = \Delta,
\label{eqn:RSFPresonance}
\end{equation}
where it is assumed that $n_n-n_e>0$.
At resonance, the transformation between
${\bar \nu_e}$ and ${\nu_\tau}$
can be greatly enhanced even if $\mu_\nu B_{\rm res} \ll \Delta$, since
the matter potential cancels the mass difference between
${\bar \nu_e}$ and ${\nu_\tau}$ at this position [see Eq.  
(\ref{eqn:RSFPresonance})]. Here
$B_{\rm res}$ is the magnitude of the transverse magnetic field at 
resonance.
At least in a mathematical sense, the RSFP level  
crossing is very similar to the MSW one.

The effective mixing angle $\tilde{\theta}$ and the
precession length $L$ in the supernova environment are given by
\begin{equation}
\sin2\tilde{\theta}
= \frac{2 \mu_\nu B}{\{(2 \mu_\nu B)^2 +[\Delta - \sqrt{2}
G_F\ (n_n-n_e)]^2\}^{1/2}},
  \label{eqn:mixinganglesin}
\end{equation}
\begin{equation}
L =
 \frac{2\pi}{\{(2 \mu_\nu B)^2 +[\Delta - \sqrt{2}
G_F\ (n_n-n_e)]^2\}^{1/2}}.
  \label{length}
\end{equation}
The precession length at the RSFP resonance is
\begin{equation}
L_{\rm res} =
 \frac{\pi}{\mu_\nu B_{\rm res}} \approx 1.1 \times 10^4\  {\mbox{cm}}
\left[\frac{\mu_{B}\cdot{\mbox{G} }}
{\mu_\nu B_{\rm res}}\right],
  \label{lengthres}
\end{equation}
where $\mu_{B}  \equiv e/2m$ is the Bohr magneton.
The width of the RSFP resonance is given by
\begin{equation}
\delta r = 2H \tan 2\tilde{\theta}_0,
  \label{width}
\end{equation}
where
\begin{eqnarray}
\tan 2\tilde{\theta}_0 \equiv & \hskip -5.4cm
{2\mu_\nu B_{\rm res}}/{\Delta} \nonumber \\
\approx  & 2.3 \times 10^{-3}
\left[\dfrac{\mu_\nu B_{\rm res}}{\mu_B\cdot{\mbox{G}}}\right]
\left[\dfrac{{\mbox{ 100 eV}}^2}{\delta m^2 }\right]
 \Biggl[ \dfrac{E_\nu}{{\mbox{ 10\  MeV}}}       \Biggr].
\end{eqnarray}
Here $H$ is the ``density'' scale height which can be expressed as
\begin{equation}
H \equiv \left| \frac{d}{dr}\ln(n_n-n_e)\right|_{\rm res}^{-1}
\approx  \left| \frac{d}{dr}\ln\rho \right|_{\rm res}^{-1}.
  \label{eqn:scaleheight}
\end{equation}
To derive Eq. (\ref{eqn:scaleheight})
we have assumed that
$|d Y_e/dr|$ is very small compared with $|d\ln\rho/dr|$.
This approximation is valid in the region above the neutrinosphere  
where RSFP can significantly affect supernova dynamics and/or  
nucleosynthesis.

Adiabatic spin-flavor conversion takes place when
the condition $L_{\rm res} \ll \delta r$ obtains.
When $\tan 2\tilde{\theta}_0 \ll 1$, we can employ the
simple Landau-Zener approximation to estimate the probability
for a $\nu_\tau$ (or $\bar \nu_e$) going through
the RSFP resonance to remain as a $\nu_\tau$ (or $\bar \nu_e$).
This probability is given by
\begin{eqnarray}
P_{\rm RSFP}
\approx& \hskip -6.2 cm
\exp\left(-\dfrac{\pi^2}{2}\dfrac{\delta r}{L_{\rm res}} \right)  
\nonumber\\
          \approx & \exp\left\{
-0.2\left[\dfrac{\mu_\nu B_{\rm res}}{\mu_B\cdot \mbox{G}}\right]^2
 \left[\dfrac{\mbox {100\ eV}^2}{\delta m^2  
}\right]
      \Biggl[ \dfrac{E_\nu}{\mbox {10\  MeV}}          \Biggr]
      \Biggl[\dfrac{H}{10^5\ \mbox {cm}  }          \Biggr]
                 \right\}.
\label{eqn:RSFPLZ}
\end{eqnarray}
It should be noted that the dependence of this probability on $E_\nu$  
and $\delta m^2$ is opposite to the corresponding depence of the
MSW survival  
probability [see Eq.  
(\ref{eqn:MSWLZ})].

It is conceivable that we may encounter situations where perhaps the  
magnetic fields are very large, or neutrino 
transition magnetic moments are  
nearly at the maximum values allowed by experiment. Such a situation,  
in turn, could lead to 
large precession effects when $\tan 2\tilde{\theta}_0 \sim 1$.
To properly treat the RSFP survival probabilities in this case, we  
should employ the following more 
appropriate formula \cite{PWH} instead of  
Eq. (\ref{eqn:RSFPLZ}),
\begin{eqnarray}
P_{\rm RSFP}
\approx\frac{1}{2}+\Biggl[\frac{1}{2}-
\exp\Biggl(-\dfrac{\pi^2}{2}\dfrac{\delta r}{L_{\rm res}} \Biggr) \Biggr]
\cos2\tilde\theta_i \cos2\tilde\theta_f
\label{eqn:RSFPLZ2},
\end{eqnarray}
where
\begin{equation}
\cos2\tilde\theta_{i(f)}
= \frac{\Delta -\sqrt{2}G_F(n_n-n_e)}{\{(2 \mu_\nu B)^2 +[\Delta -  
\sqrt{2}
G_F\ (n_n-n_e)]^2\}^{1/2}}\Bigg|_{r=r_{i(f)}}.
  \label{eqn:mixinganglecos}
\end{equation}
Here, $\tilde\theta_{i}$ is the initial mixing angle
at radius $r_i$ where neutrinos are produced (neutrinosphere), and
$\tilde\theta_{f}$ is the final mixing angle at radius $r_f$
where we calculate the survival probabilities.
The initial mixing angle at the neutrinosphere
always satisfies 
$\cos 2\tilde\theta_{i}\approx -1$
for $\mu_\nu B(r_i) \lsim 10\ \mu_B \cdot\mbox{G}$,
whereas $\cos 2\tilde\theta_{f} \approx 1$ unless
$2\mu_\nu B(r_f) \gsim \Delta $.
For our choice of parameters in Secs. III and IV,
Eqs. (\ref{eqn:RSFPLZ}) and (\ref{eqn:RSFPLZ2}) give
almost identical probabilities except
around $\delta m^2 \sim 1$ eV$^2$.

Let us now consider the case where both MSW and RSFP resonances  
occur along a neutrino trajectory. As one can see from Eq.
(\ref{eqn:resodensity}),
the resonance densities for MSW conversion and RSFP will necessarily be
different for $Y_e\neq 1/3$. This in turn implies that 
the MSW and RSFP resonances will occur at  
different distances from the neutron star.
In Fig. 2 we plot a typical matter density profile at late
times, which corresponds to $t_{\rm PB} \approx 6$ s
in a numerical supernova model by Wilson and Mayle \cite{Wilson}.
In this figure we also indicate the resonance
positions for RSFP (filled circles) and MSW conversion
(open squares) for a neutrino with $E_\nu = 25$ MeV 
and for various labeled heavier vacuum neutrino  
mass eigenvalues.
In labeling these resonance positions we have assumed that $\delta  
m^2 = m_2^2- m_1^2 \approx m_2^2$.
Note that the matter density at the neutrinosphere is about  
$10^{12}$ g cm$^{-3}$.
In Fig. 3 we plot the resonance positions (log of the radius in cm)  
for MSW conversion (dashed line) and RSFP (solid line) 
as functions of $\delta  
m^2$.
From Figs. 2 and 3, we can clearly see that for a given neutrino  
energy, the RSFP resonance occurs at higher density than the  
MSW resonace, unless $\delta m^2 > 1000$ eV$^2$.

An analytic treatment of the case where both RSFP and MSW
resonances occur along a neutrino path is possible if the two  
resonances are well separated in space. The non-overlapping condition  
for these two resonances is given by \cite{AB}
\begin{equation}
(3Y_e-1)/{Y_e}\ > \
\tan 2\theta + \tan 2\tilde{\theta}_0.
  \label{eqn:nonoverlap}
\end{equation}
If this condition is not satisfied, we would have to do a numerical
integration of Eq. (\ref{eqn:evolution1}) to estimate reliably the
survival probabilities.
The above inequality holds for almost all of the parameter region we  
will
discuss in Secs. III and IV.
The probability
for a $\nu_e$ (or $\nu_\tau$ ) going through
the MSW resonance to remain as a $\nu_e$ (or $\nu_\tau$ ) is
given by
\begin{eqnarray}
  P_{\rm MSW}\approx& \hskip -3.3cm
\exp\left(-\displaystyle {\pi\over 2}H\Delta \sin^22\theta\right) 
\nonumber \\
        \approx& \exp\left\{
-0.4 \left[\dfrac{\delta m^2}{{ \mbox{100\ eV} }^2}\right]
      \Biggl[\dfrac{\mbox{10\ MeV}}{E_\nu}     \Biggr]
      \Biggl[\dfrac{H}{10^5\ {\mbox{cm}}}  \Biggr]
       \left[\dfrac{\sin^2 2 \theta}{10^{-5}}\right]   
                 \right\},
\label{eqn:MSWLZ}
\end{eqnarray}
for $\theta\ll 1$.
In Eq. (\ref{eqn:MSWLZ}), we have made the approximation
$|d\ln n_e/dr|^{-1}\approx |d\ln\rho/dr|^{-1}\approx H$.
In Fig. 4 we plot the more accurate \lq\lq density\rq\rq\ 
scale heights for the RSFP (solid line) 
and
MSW (dashed line) resonances as 
functions of $n_n-n_e$ and $n_e$, respectively. 

By employing Eqs. (\ref{eqn:RSFPLZ2}) and (\ref{eqn:MSWLZ}),
we can easily estimate the probability that a $\nu_\tau$ emitted from
the neutrinosphere, and subsequently propagating through an RSFP  
resonance and then later through an MSW resonance, emerges as either
 $ \nu_\tau,\ \nu_e$, and ${\bar \nu_e}$ as
\begin{eqnarray}
P(\nu_\tau\rightarrow \nu_\tau)\   =& P_{\rm RSFP} P_{\rm MSW}, \\
P(\nu_\tau\rightarrow \nu_e)\      =& P_{\rm RSFP}[1- P_{\rm MSW}], \\
P(\nu_\tau\rightarrow \bar \nu_e) \  =& 1-P_{\rm RSFP}.
\end{eqnarray}
Other probabilities such as $P(\bar \nu_e \rightarrow \nu_\tau)$ and
 $P(\nu_e \rightarrow \nu_\tau)$  can be estimated in a  
straightforward and similar fashion.

\vglue 1cm
\centerline{\bf III.  Neutrino Spin-Flavor Conversion 
and Hot Bubble}
\centerline{\bf $r$-Process Nucleosynthesis}
\vglue 0.5cm

In the post-core-bounce evolution of the hot proto-neutron star, all
six species of neutrinos and antineutrinos are produced thermally,  
provided that all vacuum neutrino masses are reasonably light.
These neutrinos carry away the gravitational binding energy of the  
neutron star on a neutrino diffusion timescale. This timescale is  
roughly $\sim 10$ s and is essentially 
set by three quantities:  
the mass of the neutron star (roughly the Chandrasekhar mass), the  
saturation density of nuclear matter, and the Fermi constant $G_F$.  
The neutrino diffusion process sets the timescale for all of the  
post-core-bounce supernova evolution. We have explained in the  
introduction that we expect the neutrinos emitted from the  
neutrinosphere to be instrumental 
in heating and ejecting the envelope of  
material 
which surrounds the neutron star. For our purposes it is convenient,  
if somewhat artificial, to divide the evolution of this envelope into  
two epochs: the epoch of shock reheating at $t_{\rm PB} < 1\,{\rm  
s}$, and the hot bubble/$r$-process epoch at $t_{\rm PB}  
\approx 3$--20 s. In addition to the neutrino diffusion  
timescale, the neutron star radius, the weak and  
nuclear freeze-out positions, and the location of the rapid neutron  
capture environment, there is yet one more important characteristic  
length scale in the hot bubble/wind environment --- the gain radius. 
Heating engendered by charged-current absorption of electron  
neutrinos and antineutrinos on nucleons wins out over 
neutrino losses in the hot plasma above the \lq\lq gain  
radius,\rq\rq\ 
 $r_g \gsim 10 $ km
\cite{QW96,Bethe}. 

As outlined in Sec. I, early deleptonization of the hot  
proto-neutron star causes its outer layers to become  
neutron rich. This neutron excess causes
$\nu_e$ to have a larger opacity than 
$\bar \nu_e$ because of
the charged-current capture reactions on free nucleons in
Eqs. (\ref{eqn:beta1a}) and (\ref{eqn:beta1b}).
Consequently, $\bar \nu_e$ have a larger average energy than 
$\nu_e$ because $\bar \nu_e$ decouple deeper in the core where  
the matter is hotter.
The typical average energies for $\nu_e$ and $\bar \nu_e$ at the
hot bubble/$r$-process epoch are 11 and 16 MeV, respectively.
On the other hand, the typical average energies of
$\nu_\mu$, ${\bar\nu_\mu}$, $\nu_\tau$ and ${\bar\nu_\tau}$
are all about 25 MeV.
This is because they only have the neutral-current reaction opacity  
sources common to all neutrino species. 
Hence, $\nu_\mu$, ${\bar\nu_\mu}$, $\nu_\tau$ and  
${\bar\nu_\tau}$ 
decouple in regions hotter than those where
$\nu_e$ and $\bar \nu_e$ decouple.
Thus, the average neutrino energies 
in supernovae during the post-core-bounce epoch 
always satisfy the hierarchy:
\begin{equation}
\langle E_{ \nu_{\tau(\mu)}}\rangle\approx
\langle E_{\bar \nu_{\tau(\mu)}}\rangle
>\langle E_{\bar \nu_e}\rangle
>\langle E_{\nu_e}\rangle.
  \label{eqn:averageenergies}
\end{equation}

The value of $Y_e$ in the region above the neutrinosphere is
determined by the charged-current reactions in Eqs.
(\ref{eqn:beta1a}) and (\ref{eqn:beta1b}).
The rate of change of $Y_e$
with time $t$ or radius $r$ in this region is given by
\begin{equation}
\frac{dY_e}{dt}=v(r)\frac{dY_e}{dr}=
\lambda_1-\lambda_2 Y_e,
  \label{eqn:electronfraction}
\end{equation}
where $v(r)$ is the radial velocity field of the material in the  
supernova.
In this equation, $\lambda_1=\lambda_{\nu_e n}+\lambda_{e^+ n}$ and
$\lambda_2=\lambda_1 +\lambda_{{\bar \nu_e} p}+\lambda_{e^- p}$.
Here, $\lambda_{\nu_e n}$ and $\lambda_{{\bar \nu_e} p}$
denote the rates of the reactions in Eqs. (\ref{eqn:beta1a}) and
(\ref{eqn:beta1b}),
respectively, and
$\lambda_{e^- p}$ and $\lambda_{e^+ n}$ denote the rates
for their reverse reactions.
At some point above the neutrinosphere the local material
expansion rate in the hot bubble 
will be faster than the rates of the reactions in Eqs.
(\ref{eqn:beta1a}) and (\ref{eqn:beta1b}).
We shall term this location the weak freeze-out point, since the  
value of
$Y_e$ for  material flow above this point will remain constant
in time and space.
Above the weak freeze-out point the solution of Eq.  
(\ref{eqn:electronfraction})
gives
\begin{equation}
Y_e(r_{\rm NFO})\approx Y_e(r_{\rm WFO})
\approx
\frac{1}
{1+\lambda_{{\bar \nu_e} p}(r_{\rm WFO})/\lambda_{{\nu_e}  
n}(r_{\rm WFO})},
\label{eqn:electronfraction2}
\end{equation}
where $r_{\rm NFO}$ and $r_{\rm WFO}$ are the nuclear and
weak freeze-out radius, respectively.
Here we have neglected $\lambda_{e^- p}$ and $\lambda_{e^+ n}$.
This is a valid approximation for our purposes, since the matter  
temperature in the region above the gain radius is small compared 
with the
effective temperatures for $\nu_e$ and $\bar \nu_e$ 
energy distributions,
and hence they are small compared with
$\lambda_{{\bar \nu_e} p}$ and $\lambda_{{\nu_e} n}$.

The rate $\lambda_{\nu N}$ can be calculated as
\begin{equation}
\lambda_{\nu N} \approx \frac{L_\nu}{4 \pi r^2}
\frac{\displaystyle \int_0^\infty \hskip -0.3cm \sigma_{\nu  
N}(E_\nu)
f_{\nu}(E_\nu)dE_\nu}
{\displaystyle \int_0^\infty \hskip -0.3cm E_\nu f_{\nu}(E_\nu)dE_\nu},
\label{eqn:lambda}
\end{equation}
with $(\nu,\ N)=(\nu_e,\ n)$ or $(\bar\nu_e,\ p)$.
Here, $L_\nu$ is the neutrino luminosity, $f_\nu(E_\nu)$ is the  
normalized neutrino
energy distribution function. We take $f_\nu(E_\nu)$ to be  
Fermi-Dirac with zero chemical potential in character
for all neutrino species, i.e.,
\begin{equation}
f_\nu(E_\nu) =
\frac{1}{1.803}\frac{1}{T_\nu^3}
\frac{E_\nu^2}
{1+\exp[E_\nu/T_\nu]},
\label{eqn:FD}
\end{equation}
where $T_\nu$ is the neutrino temperature.
The temperatures characterizing the distribution functions of each  
neutrino species at late
epochs are approximately given by
$T_{\nu_e} \approx 3.5$ MeV, $T_{\bar \nu_e} \approx 5.1$ MeV,
and $T_{\nu_{\tau(\mu)}}\approx T_{\bar \nu_{\tau(\mu)}}
\approx 7.9$ MeV. In Eq. (\ref{eqn:lambda}),
$\sigma_{\nu N}$ is the cross section for the reactions in Eqs.
(\ref{eqn:beta1a}) and (\ref{eqn:beta1b}), and is approximately
given by
\begin{equation}
\sigma_{\nu N} \approx 9.6 \times 10^{-44}
\left(\frac{E_\nu}{\rm MeV}\right)^2
\ {\mbox{cm}}^2.
 \label{eqn:sigma}
\end{equation}

Utilizing Eqs. (\ref{eqn:lambda})--(\ref{eqn:sigma}), 
and taking into account that
each neutrino species has roughly the same luminosity at late epochs,
we can approximate Eq. (\ref{eqn:electronfraction2}) as
\begin{equation}
Y_e(r_{\rm NFO}) \approx \frac{1}{1+\langle E_{\bar \nu_e}\rangle
/\langle E_{\nu_e}\rangle}
\approx \frac{1}{1+T_{\bar \nu_e}/ T_{\nu_e}}.
\end{equation}
For typical neutrino average energies
$\langle E_{\bar \nu_e}\rangle =16$ MeV and
$\langle E_{\nu_e}\rangle = 11$ MeV, we obtain $Y_e \approx 0.41$.
This value can be regarded as the standard
supernova model prediction for the $Y_e$ in the hot bubble
in the absence of RSFP and MSW conversion. 

Let us examine now how the value of $Y_e$ would be affected
by RSFP and MSW resonances occurring along neutrino trajectories
below the weak freeze-out radius in the hot bubble.  
For this case, the computation of $Y_e$ must employ the \lq\lq  
distorted\rq\rq\ energy distribution
functions for $\bar \nu_e$ and $\nu_e$ which will result from the  
energy-dependent flavor conversion associated with the neutrino  
propagating through an RSFP and/or an MSW resonance.
By using the survival probabilities calculated in Eqs.
(\ref{eqn:RSFPLZ}) and (\ref{eqn:MSWLZ}) 
in Sec. II at these RSFP and MSW  
resonances, we can estimate the
effective $\bar\nu_e$ and $\nu_e$ energy distribution functions  
at the 
weak freeze-out radius to be
\begin{equation}
f_{\bar \nu_e}(E_\nu) =
f_{\bar \nu_e}^0(E_\nu) P_{\rm RSFP}(E_\nu)
+  f_{\nu_\tau}^0(E_\nu)
[1- P_{\rm RSFP}(E_\nu)],
\label{eqn:distribution1}
\end{equation}
\begin{eqnarray}
f_{\nu_e}(E_\nu) =
& 
f_{\nu_e}^0(E_\nu) P_{\rm MSW}(E_\nu) \nonumber\\
&+ f_{\bar \nu_e}^0(E_\nu)
 [1-P_{\rm RSFP}(E_\nu)]
[1-P_{\rm MSW}(E_\nu)]\nonumber \\
& f_{\nu_\tau}^0(E_\nu)P_{\rm RSFP}(E_\nu)[1- P_{\rm MSW}(E_\nu)].
\label{eqn:distribution2}
\end{eqnarray}
Here $f_\nu^0(E_\nu)$ 
represents the appropriate initial neutrino energy distribution  
function. As in Eq. (\ref{eqn:FD}), these 
initial neutrino energy distribution  
functions are all
assumed to be Fermi-Dirac with zero chemical potential
in character, but with different
temperatures $T_{\nu_e}$, $T_{\bar \nu_e}$ and $T_{\nu_\tau}$.
By employing the distorted energy distribution functions for  
$\bar\nu_e$ and
$\nu_e$ in Eqs. (\ref{eqn:distribution1}) and  
(\ref{eqn:distribution2}), respectively,
we can use
Eq. (\ref{eqn:lambda}) to calculate the rate $\lambda_{\nu N}$, and  
hence Eq. (\ref{eqn:electronfraction2}) to estimate $Y_e$.

We can now describe the results of our calculation of $Y_e$
for  two cases: (1) RSFP but no MSW conversion along a  
neutrino trajectory, and (2) both RSFP and MSW conversion on the same  
neutrino path.
In our calculations we assume that the
magnetic field profiles around the proto-neutron star are as follows:
\begin{equation}
B(r)=B_0 \ (r_0/r)^n\times 10^{12}\  \mbox{G},
\label{eqn:profile}
\end{equation}
where $r_0 = 10$ km and $n$ = 2 or 3.
A magnetic field of order $10^{12}$ G
around the proto-neutron star is plausible,  
especially given that some pulsar magnetic fields are at least this  
large. However, this argument should be viewed with some skepticism.  
There is no guarantee, for example, that the large pulsar magnetic  
fields are not generated only after the epoch of neutrino heating  
that we are investigating. Nevertheless, the RSFP effects which we  
describe actually depend on the product of the neutrino transition  
magnetic moment and the magnetic field component transverse to the  
neutrino trajectory. So in addition to any uncertainties in the  
magnetic field magnitude, there are geometric uncertainties due to  
the unknown distribution and orientation of the magnetic field, as  
well as the inherent uncertainties in the neutrino transition magnetic  
moment.
For the purposes of our parametric study, we fix the value of the  
neutrino transition magnetic moment to be
$\mu_\nu= 
10^{-12} \mu_B$. Again, however, it should
be noted that our results depend only on the
combination of $\mu_\nu B_0$ for fixed value of $n$.

Let us first discuss case (1) where there is only RSFP
and no MSW conversion. This case will obtain whenever
$\sin^22\theta \ll 10^{-5}$ and $\delta m^2 \approx 1$--$10^4$ eV$^2$.
We use the energy distributions in Eqs.
(\ref{eqn:distribution1}) and (\ref{eqn:distribution2}) with
$P_{\rm MSW} = 1$ to calculate the values of $Y_e$ in this case.
The initial and final mixing angles for RSFP in this case are taken  
to be
at the neutrinosphere ($r\approx 10$ km) and the weak freeze-out  
radius ($r\approx
40$ km),
respectively.
In Fig. 5 we have plotted
the contours of $Y_e$ (as labeled) in the $B_0$-$\delta m^2$ plane
for $B\propto r^{-2}\ (n = 2)$. These contours  
correspond to $Y_e$ values at the weak freeze-out radius,
above which $r$-process nucleosynthesis  
may be occurring.
One can see from this figure that the value of $Y_e$ is
modified significantly from the standard no-RSFP case whenever $B_0  
\gsim 0.1$.
In Fig. 6 we have plotted the same contours as in Fig. 7,
but for $B\propto r^{-3}\ (n = 3)$. Because of the rapid decrease
of magnetic field for $n = 3$,  
the magnitude of $B_0$ required to cause similar effects on $Y_e$ is
somewhat larger than that for $n = 2$.

Next we examine case (2), where there exist both
RSFP and MSW conversion along neutrino trajectories.
In Fig. 7 we plot the regions in the
$\sin^2 2\theta$-$\delta m^2$ parameter space where   
neutrino flavor evolution is dominated by RSFP and/or MSW conversion,
or neither, for the matter density profile in Fig. 2.
The solid and dashed lines correspond to $P_{\rm RSFP}=e^{-1}$ and
$P_{\rm MSW}=e^{-1}$, respectively, for a neutrino with
$E_\nu$ = 25 MeV. We have taken 
$\mu_\nu B_{\rm res}=1\ \mu_B\cdot\mbox{G}$
in drawing the solid line in Fig. 7. This figure
clearly separates the parameter space into four regions, (a)--(d).
In region (a), MSW conversion
is nearly completely adiabatic and therefore very  
efficient in $\nu_e\leftrightarrow\nu_\tau$ conversion; whereas
in region (b), RSFP is nearly completely adiabatic and thus quite  
efficient in $\bar\nu_e\leftrightarrow\nu_\tau$ conversion.
In region (c), neutrino propagation is quite adiabatic through both  
the RSFP and MSW resonances; whereas in region (d), neutrino  
propagation through either type of resonance is
non-adiabatic.
The solid line in this figure would move upward to
$\delta m^2 \approx 1000$ eV$^2$ if we were to take 
$\mu_\nu B_{\rm res}=10\ \mu_B\cdot\mbox{G}$, and   
downward to
$\delta m^2 \approx 3$ eV if we were to take 
$\mu_\nu B_{\rm res}=0.1\ \mu_B\cdot\mbox{G}$.

In the calculations for case (2), the initial mixing angle (required for  
estimating survival probabilities after neutrino propagation through  
an RSFP resonance) is evaluated at
the neutrinosphere. This choice is not crucial to our results  
--- it is only necessary that we evaluate the initial mixing angle in a  
region where the density is significantly larger than that at the  
resonance position. The final mixing angle  
$\tilde\theta_{f}$
is also required for estimating RSFP survival probabilities.  
This angle must be evaluated at a radius lying somewhere above the  
RSFP resonance,
but below the MSW resonance.
We have used the following approximation to estimate  
$\tilde\theta_{f}$:
\begin{equation}
\cos 2\tilde\theta_{f} = \left\{
\begin{array}{@{\,}ll}
\cos 2\tilde{\theta}|_{r=r_{\rm MSW}}
&\ \    {\rm if}\ \  r_{\rm MSW} < r_{\rm WFO},\\
\cos 2\tilde{\theta}|_{r=r_{\rm WFO}}
&\ \    {\rm if}\ \  r_{\rm MSW} > r_{\rm WFO},\\
\end{array}
\right.
\end{equation}
where $r_{\rm MSW}$ is the radius for the MSW resonance.
The non-overlapping condition
for the RSFP and MSW resonances in Eq. (\ref{eqn:nonoverlap}) described  
in Sec. II
must be satisfied if we are to employ the analytic Landau-Zener  
formula for survival probabilities.
Since the two resonances are well separated
so long as $\delta m^2 \lsim 1000$ eV$^2$, we choose to
perform calculations for
this parameter range only.

In Fig. 8 we have plotted contours of $Y_e$ (as labeled) in the
$\sin^22\theta$-$\delta m^2$
plane for 
the magnetic field profile of the form given in Eq.  
(\ref{eqn:profile}) with 
$n=2$ and $B_0$ = (a) 0.01,  
(b) 0.1, (c) 1.0, and (d) 10.0 on separate plots.
In plots (a) and (b), the magnetic field
is too small to cause appreciable RSFP, and
thus we have obtained essentially the
same results as in Ref. \cite{QFMMW}. 
However, in plots (c) and (d), the contours corresponding to  
$Y_e = 0.5$ are
significantly altered over the case without RSFP. In particular,   
in plot (d) where $B_0=10.0$,
there is {\it no} parameter region where $Y_e > 0.5 $.
We have also done a similar set of calculations but now with  
$n=3$
and $B_0$ = (a) 0.1, (b) 0.5, (c) 1.0, and (d) 10.0. Contour  
plots for these calculations are given in Fig. 9. The overall  
qualitative behavior exhibited in Fig. 9 is similar to that
in Fig. 8.
From both Figs. 8 and 9 we can conclude that the effects of RSFP  
will dominate over those of MSW conversion alone whenever
$B_0 \gsim 1.0$. Of course, this particular {\it quantitative}  
conclusion is predicated on neutrino transition magnetic moments  
being near their maximally allowed values. Obviously, smaller  
neutrino transition magnetic moments would then necessitate a larger  
threshold value of $B_0$ for which RSFP would dominate 
the effects on $Y_e$ in  
the hot bubble.

\vglue 1cm
\centerline{\bf IV. Neutrino Spin-Flavor Conversion in 
  the Shock Reheating Epoch}
\vglue 0.5cm
Neutrino propagation through RSFP resonances in the region above the  
neutrinosphere can also affect the dynamics of the supernova
explosion. This comes about because flavor  
conversion in the region below the shock during the  
reheating epoch, $t_{\rm PB} \approx 0.1$--0.6 s, 
can lead to an enhanced  
neutrino energy deposition rate in this region which, in turn, can  
lead to a higher shock energy.
For example, $\bar \nu_e$ in the reheating region  
would become
more energetic if high energy $\nu_\mu$ or $\nu_\tau$ propagate through  
an RSFP resonace and become $\bar\nu_e$. This would lead to an  
enhanced $\bar\nu_e$ capture rate on protons, and hence an enhanced  
heating rate. Therefore, in what follows, 
we concentrate on how RSFP and/or MSW  
resonances influence the charged-current neutrino and antineutrino  
heating rates in the region below the shock.

The shock heating rate per proton or neutron is given by
\begin{equation}
  \dot\epsilon_{\nu N}
  \approx \frac{L_\nu}{4\pi r^2}
  \frac{\int_0^\infty E_\nu f_\nu (E_\nu) \sigma_{\nu N}dE_\nu }
  {\int_0^\infty E_\nu f_\nu (E_\nu)dE_\nu}.
\label{eqn:heat}
\end{equation}
The total heating rate accompanying the $\nu_e$ and $\bar\nu_e$  
absorption processes in
Eqs. (\ref{eqn:beta1a}) and  
(\ref{eqn:beta1b}) is
given by
\begin{equation}
  \dot\epsilon_{\rm tot}
=Y_n   \dot\epsilon_{\nu_e n}+Y_p   \dot\epsilon_{{\bar \nu_e}  
p},
\end{equation}
where $Y_n$ and $Y_p$ are the number fractions of
free neutrons and protons, respectively, and are approximately
specified by Eq. (\ref{eqn:electronfraction2}) as $Y_n\approx 1-Y_e$ and
$Y_p\approx Y_e$, respectively.
The ratio of the total heating rate with RSFP (primed symbols) to  
that without RSFP (and without MSW conversion) is given by
\begin{equation}
  \frac{\dot\epsilon_{\rm tot}^\prime}{\dot\epsilon_{\rm tot}}
\approx
\frac{Y_n^\prime   \dot\epsilon_{\nu_e n}^\prime+Y_p^\prime
   \dot\epsilon_{{\bar  
\nu_e} p}^\prime}
{Y_n   \dot\epsilon_{\nu_e n}+Y_p   \dot\epsilon_{{\bar \nu_e}  
p}}.
\label{eqn:reheat1}
\end{equation}

At a representative time during the reheating epoch, the  
temperatures of the relevant neutrino species are approximately given  
by
$T_{\nu_{\tau(\mu)}}=T_{\bar\nu_{\tau(\mu)}}\approx 7$ MeV and
$T_{\nu_e}\approx T_{\bar \nu_e}\approx 5$ MeV in the Wilson and  
Mayle calculations \cite{Wilson}.
Thus, the ratio of the total heating rates with and without RSFP  
effects can be estimated to be
\begin{equation}
  \frac{\dot\epsilon_{\rm tot}^\prime}{\dot\epsilon_{\rm tot}}
\approx
Y_n^\prime\left[1+
\frac{Y_p^\prime}{Y_n^\prime}\left(\frac{T_{\nu_\tau}}{T_{\nu_e}}\right)^2
\right]
\approx \frac{T_{\nu_\tau}}{T_{\nu_e}}
\approx 1.4,
\label{eqn:reheat2}
\end{equation}
where again we have taken the individual neutrino luminosities
to be approximately the same.
This estimate of the heating enhancement factor for RSFP has assumed  
complete $\nu_\tau \leftrightarrow
{\bar \nu_e}$ conversion in the region 
below the shock. Because $\nu_e$ 
and $\bar\nu_e$ have roughly the same energy distributions at the
reheating epoch, the heating enhancement factor and the accompanying
$Y_e$ value would remain essentially unchanged if additional MSW
$\nu_\tau\leftrightarrow \nu_e$ conversion were to follow the
complete $\nu_\tau\leftrightarrow\bar\nu_e$ conversion by RSFP.
We should note that the absolute average  
energies and therefore the temperatures of the various neutrino  
species are a subject of great debate and controversy in the  
numerical supernova modeling community. However, it is clear that 
only the {\it differences} between the temperatures of the relevant  
neutrino species are important in our estimates of the reheating  
enhancement. Neutrino transport calculation estimates of these  
differences are somewhat more reliable than those of  
the average neutrino energies themselves.   

In Fig. 10 we plot an example matter density profile in the region  
above the neutron star
at an early epoch, $t_{\rm PB}= 0.15 $ s, when shock reheating has  
commenced \cite{Wilson}.
At this time in the Wilson and Mayle model the shock wave is located  
at
$r \approx 4.7 \times 10^{7}$
cm from the neutron star center.
Here we employ the same
power-law type magnetic field profiles as in Sec. III for our  
estimates of the shock reheating enhancement, i.e.,
\begin{equation}
B(r)=B_1 \ (r_1/r)^n\times 10^{12}\  \mbox{G},
\label{eqn:profile2}
\end{equation}
where $r_1 = 100$ km and $n$ = 2 or 3.

In Fig. 11 we plot the regions in the $B_1$-$\delta m^2$  
parameter space where
the enhancement of the reheating rate is 40\%.
Contours in this  
figure are shown for $n$ = 2 and 3, as represented by the
solid and dashed lines, respectively, and are calculated from
Eq. (\ref{eqn:reheat1}) with the use of 
Eqs. (\ref{eqn:electronfraction2}), 
(\ref{eqn:lambda}),
(\ref{eqn:distribution1}), and (\ref{eqn:heat}).  If the
parameters $\delta m^2$ and $B_1$ fall inside these 
contour lines, then the  
increase in the neutrino heating rate in the region below
and near the shock
is 40\% more than that in the  
standard case without neutrino flavor mixing.
We conclude that RSFP can produce heating effects which are similar  
to those
discussed in Ref. \cite{FMMW} for the MSW conversion.
Therefore, RSFP may be beneficial to models of the supernova  
explosion, by virtue of
increasing the average $\bar\nu_e$ energy 
and leading to a more energetic shock wave. However, we should note  
that the
larger neutron-to-proton ratio (lower $Y_e$) necessarily resulting  
from RSFP
(as detailed in Sec. III) would aggravate the problem  
of
overproduction of the $N=50$ nuclei,
particularly, $^{88}$Sr, $^{89}$Y, and $^{90}$Zr, which is inherent in  
some models of the nucleosynthesis from neutrino-heated supernova
ejecta in  
this epoch \cite{Woosley3}.

\vglue 0.5cm
\centerline{\bf V. Conclusions}
\vglue 0.5cm
We have investigated the combined effects of
matter-enhanced MSW conversion and RSFP of Majorana  
neutrinos on supernova dynamics and heavy element
nucleosynthesis in neutrino-heated supernova ejecta. 
If neutrinos are of Majorana type and have transition magnetic  
moments
of order $\sim 10^{-12}\mu_B$, then in the presence of a neutron
star magnetic field of order $\sim10^{12}$ G,
resonant spin-flavor precession (conversion)
of high energy $\nu_\tau$ (or $\nu_\mu$) with vacuum masses 
in the range 1--100 eV into light $\bar \nu_e$ could increase the
neutron-to-seed ratio for the $r$-process. In principle,  
RSFP could fix the central problem which confounds current models of  
$r$-process nucleosynthesis
from late-time neutrino-heated supernova ejecta 
--- obtaining a high enough neutron-to-seed ratio. We have found that  
significant enhancement in this ratio could be obtained for a  
plausible range of the proto-neutron star
magnetic field,
$B\gsim 10^{12}$ G, and for neutrino transition  
magnetic moments near the maximum value allowed by 
the stringent astrophysical constraint.
Although RSFP effects may enable $r$-process nucleosynthesis to  
proceed in the hot bubble, it does so at a price. The enhancement of  
the $\bar\nu_e$ energies resulting from RSFP may be at odds with the  
observations of these neutrinos from SN1987A \cite{FQW96}.
Of course, on the other hand,
these considerations of the RSFP effects in supernovae
may have important implications for cosmology, since the  
range of neutrino masses, 1--100 eV, required to obtain RSFP in the  
relevant region of the supernova
is coincidently the range of interest for a significant neutrino dark  
matter component.

We have also shown here that
$r$-process nucleosynthesis would not be suppressed even if
the MSW resonance occurs along with the RSFP. This evasion of  
the bounds \cite{QFMMW} from the $r$-process on the MSW conversion
will be operative so  
long as we can be guaranteed a large
value of $\mu_\nu B_{\rm res}\gsim 1\ \mu_B \cdot\mbox{G}$.
For some of the parameters we considered,
the value of $Y_e$ can be as small as 0.3 when RSFP for  
high energy neutrinos occurs near the 
weak freeze-out point. These very low values of $Y_e$ are produced in  
our calculations when RSFP dominates neutrino flavor  
evolution. The existence of MSW resonances is irrelevant in this  
case.

We also examined the effects of RSFP during
the shock reheating epoch. We found that RSFP can increase the
total reheating rate by about 40\%, but
at the same time the concomitant reduction of $Y_e$ would exacerbate  
the problem of
the overproduction of the $N=50$ nuclei at this epoch.
This increase in total reheating rate {\it may be} welcome for
the delayed supernova explosion mechanism, which
relies on the energy deposited by $\nu_e$ and
$\bar \nu_e$ absorption reactions above the neutrinosphere to power  
the shock.
On the other hand, there is at present no compelling necessity in  
supernova models for an added boost in shock energy from a scheme
such as the RSFP of supernova neutrinos. For example, Wilson and
Mayle \cite{Wilson} obtained a supernova explosion energy in 
agreement with the SN1987A observation by the delayed mechanism
with ordinary neutrino heating alone.

\vglue 1cm
\centerline{\bf Acknowledgement}
We want to thank J. R. Wilson and R. W. Mayle
for providing us with the output from their
numerical supernova models.
We would like to acknowledge the
Institute for Nuclear Theory at University of
Washington, Seattle for its hospitality during the time
the main part of this work was done.
We would also like to thank A. Yu. Smirnov,
S. Petcov, M. Kawasaki, and A. Rossi 
for many helpful discussions. This work was supported  
in part by NSF Grant No. PHY95-03384 and NASA Grant No. NAG5-3062 at UCSD.  
H. Nunokawa was supported by a DGICYT postdoctoral fellowship at
Universitat de Val\`encia.
Y.-Z. Qian was supported by the D. W. Morrisroe Fellowship at  
Caltech.
\vfill
\eject

\vfill
\eject
\centerline{\bf Figure Captions}
\vglue 0.5cm
\noindent
Fig. 1: Schematic picture of the RSFP and MSW resonances
for two generations of Majorana neutrinos.
Each curve shows the effective neutrino mass-squared difference
as a function of the matter density
in neutron-rich ($1/3<Y_e<1/2$) matter in the presence of a magnetic
field,
including effects of a neutrino vacuum mixing angle
and a neutrino transition magnetic
moment.
\vglue 0.5cm
\noindent
Fig. 2: A typical matter density profile from numerical supernova models
at late times ($t_{\rm PB}= 5.8$ s). Filled circles and
open squares show the positions
of RSFP and MSW resonances, respectively, for a neutrino
with $E_\nu=25$ MeV and for the cases where the
heavier
vacuum neutrino masses are 100, 30, 10, 5, and 1 eV.
Numbers with \lq\lq eV\rq\rq\ are for RSFP resonances,
whereas numbers without ``eV'' are for MSW resonances.
\vglue 0.5cm
\noindent
Fig. 3: Positions for RSFP (solid line) and MSW (dashed line)
resonances for
different values of $\delta m^2$. These resonance positions
are for a neutrino with
$E_\nu=25$ MeV and correspond to the matter density profile in
Fig. 2.
\vglue 0.5cm
\noindent
Fig. 4: \lq\lq Density\rq\rq\  scale heights $H$
as functions of $n_n-n_e$ (RSFP) and $n_e$ (MSW)
corresponding to the matter density profile in Fig. 2.
\vglue 0.5cm
\noindent
Fig. 5: Contour plot of $Y_e$ in the $B_0$-$\delta m^2$
plane for the RSFP case where $B(r)=B_0(r_0/r)^2\times 10^{12}$ G
($r_0=10$ km).
\vglue 0.5cm
\noindent
Fig. 6: Same as Fig. 5 but for the case where
$B(r)=B_0(r_0/r)^3 \times 10^{12}$ G.
\vglue 0.5cm
\noindent
Fig. 7: Regions of the $\sin^2 2\theta $-$\delta m^2$ parameter
space where neutrino flavor
evolution is dominated by RSFP and/or MSW conversion, or neither,
for the matter density profile in Fig. 2.
The solid and dashed lines correspond to $P_{\rm RSFP}=e^{-1}$ and
$P_{\rm MSW}=e^{-1}$, respectively, for a neutrino with
$E_\nu$ = 25 MeV. We take $\mu_\nu B_{\rm res}=1\ \mu_B\cdot{\rm G}$
in computing the location of the solid line.
\vglue 0.5cm
\noindent
Fig. 8: Contour plots of $Y_e$ in the $\sin^22\theta$-$\delta m^2$
plane for the cases where $B(r)=B_0(r_0/r)^2\times 10^{12}$ G
($r_0=10$ km) with $B_0$ =
(a) 0.01, (b) 0.1, (c) 1.0, and (d) 10.0.
\vglue 0.5cm
\vfill
\eject
\noindent
Fig. 9: Same as Fig. 8 but for the cases where
$B(r)=B_0(r_0/r)^3\times 10^{12}$ G with $B_0=$
(a) 0.1, (b) 0.5, (c) 1.0, and (d) 10.0.
\vglue 0.5cm
\noindent
Fig. 10: A matter density profile from numerical supernova models
at an early epoch ($t_{\rm PB}= 0.15$ s).
\vglue 0.5cm
\noindent
Fig. 11: Regions of the $B_1$-$\delta m^2$ parameter space
where the increase in the total reheating
rate is 40\%. 
The solid and dashed lines are for the cases
where $B(r)=B_1(r_1/r)^n\times 10^{12}$ G ($r_1=100$ km)
with $n=2$ and 3, respectively.
\vglue 0.5cm
\vfill
\eject
\newpage
\vglue 2cm
\centerline{
\psfig{file=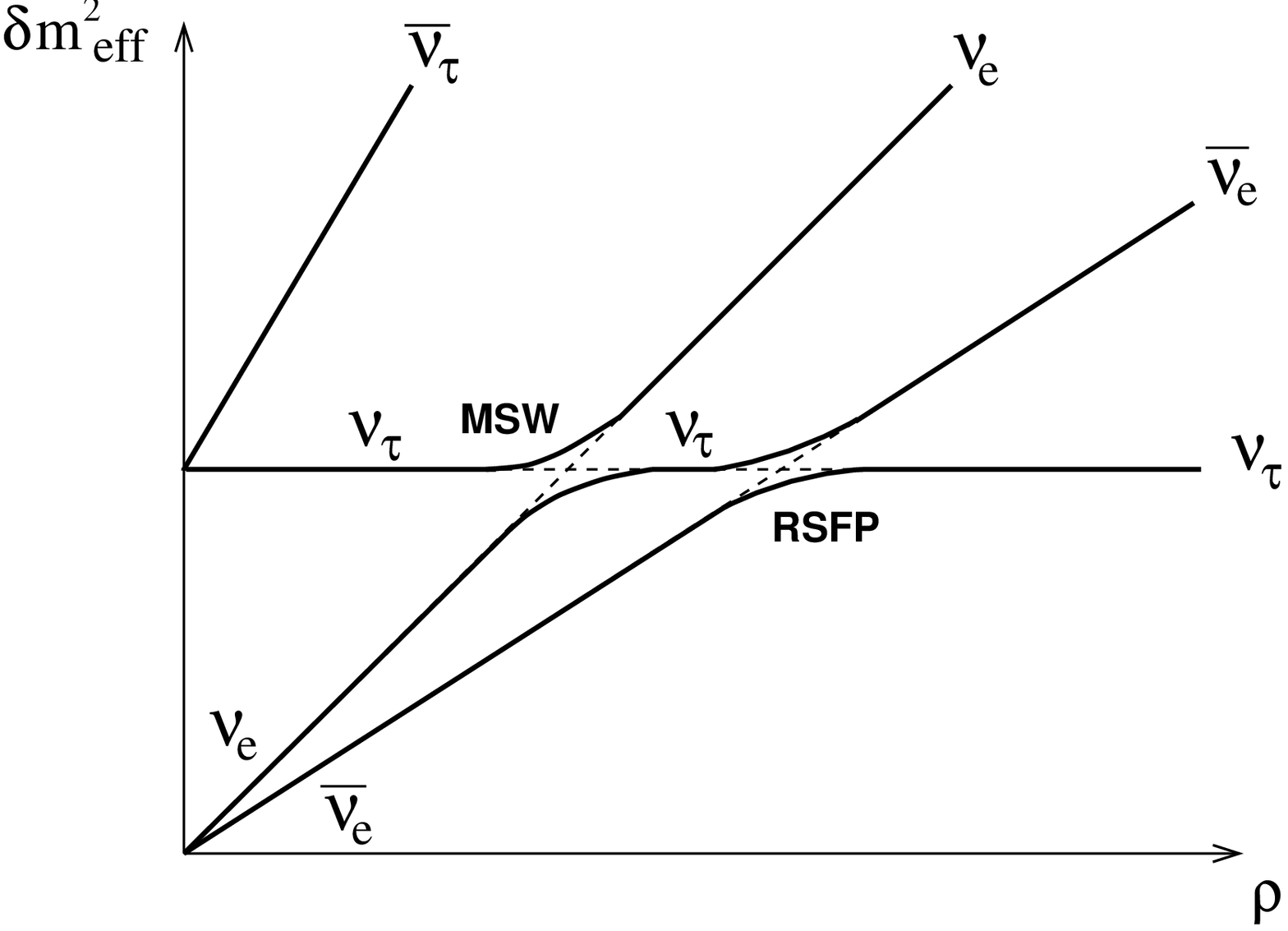,height=14.0cm,width=14.0cm,angle=-90}}
\vglue 2cm
\bf \Huge Fig. 1
\newpage
\centerline{
\psfig{file=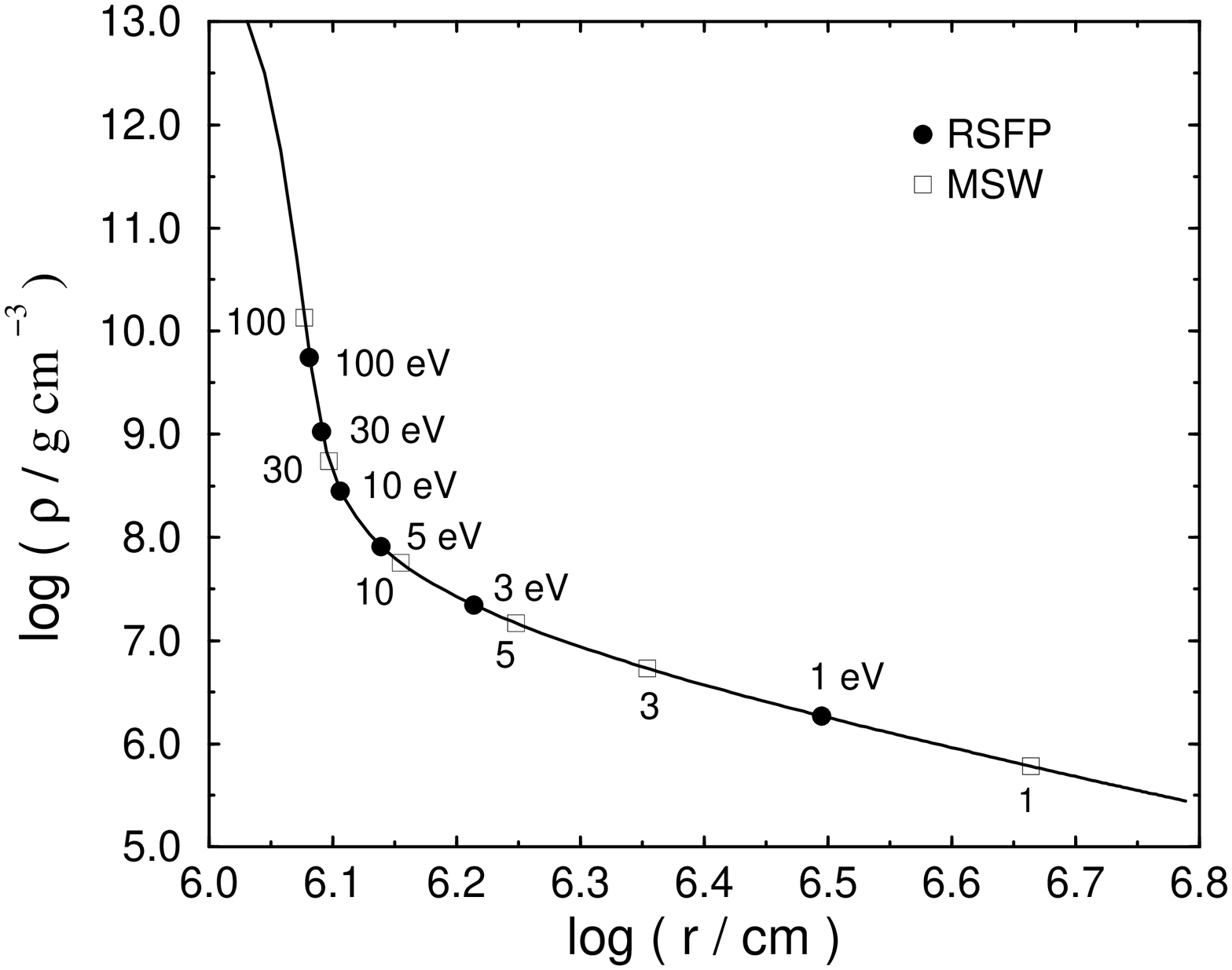,height=16.0cm,width=16.0cm,angle=-90}}
\vglue 2cm
\bf \Huge Fig. 2
\newpage
\centerline{
\psfig{file=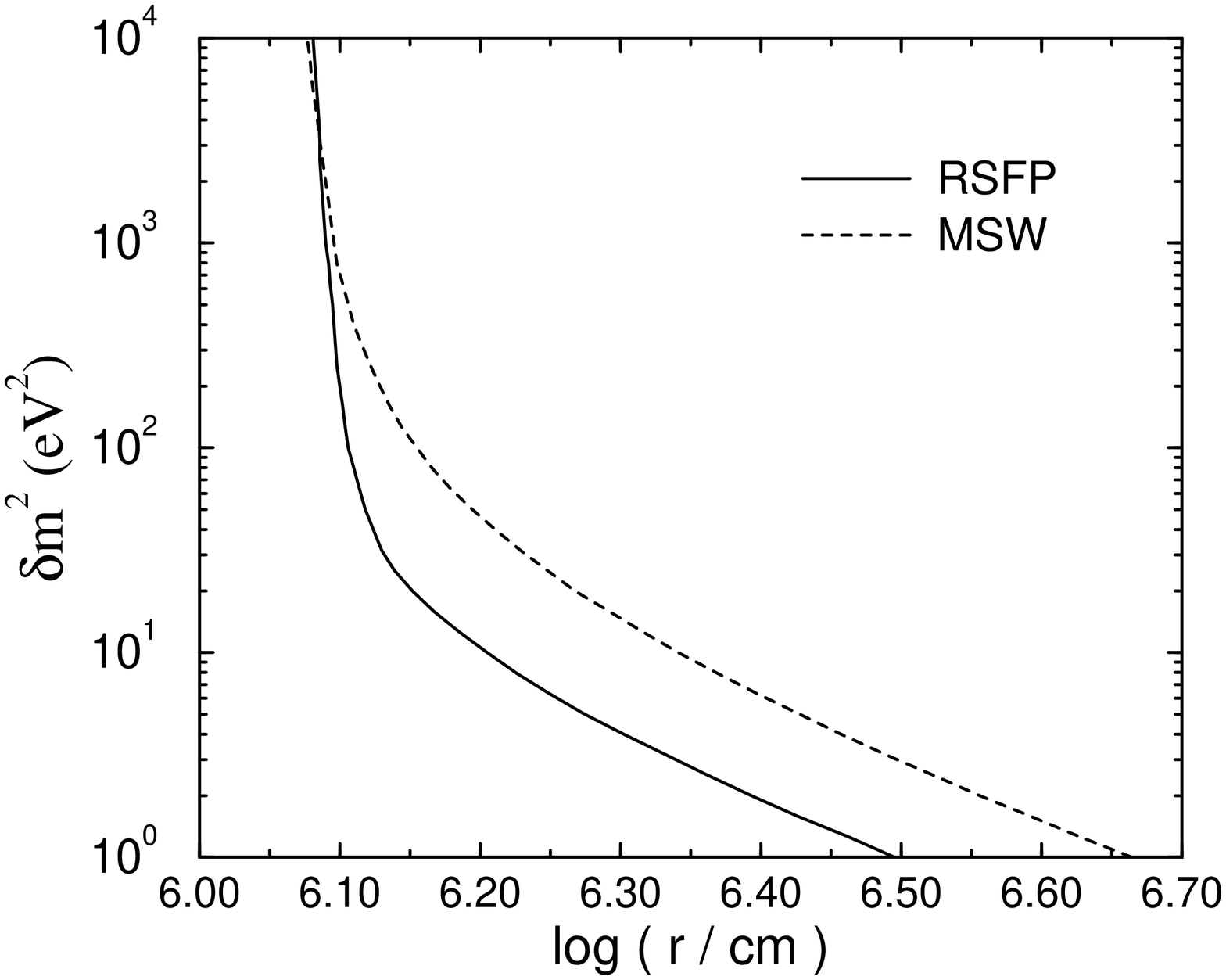,height=16.0cm,width=16.0cm,angle=-90}}
\vglue 2cm
\bf \Huge Fig. 3
\newpage
\centerline{
\psfig{file=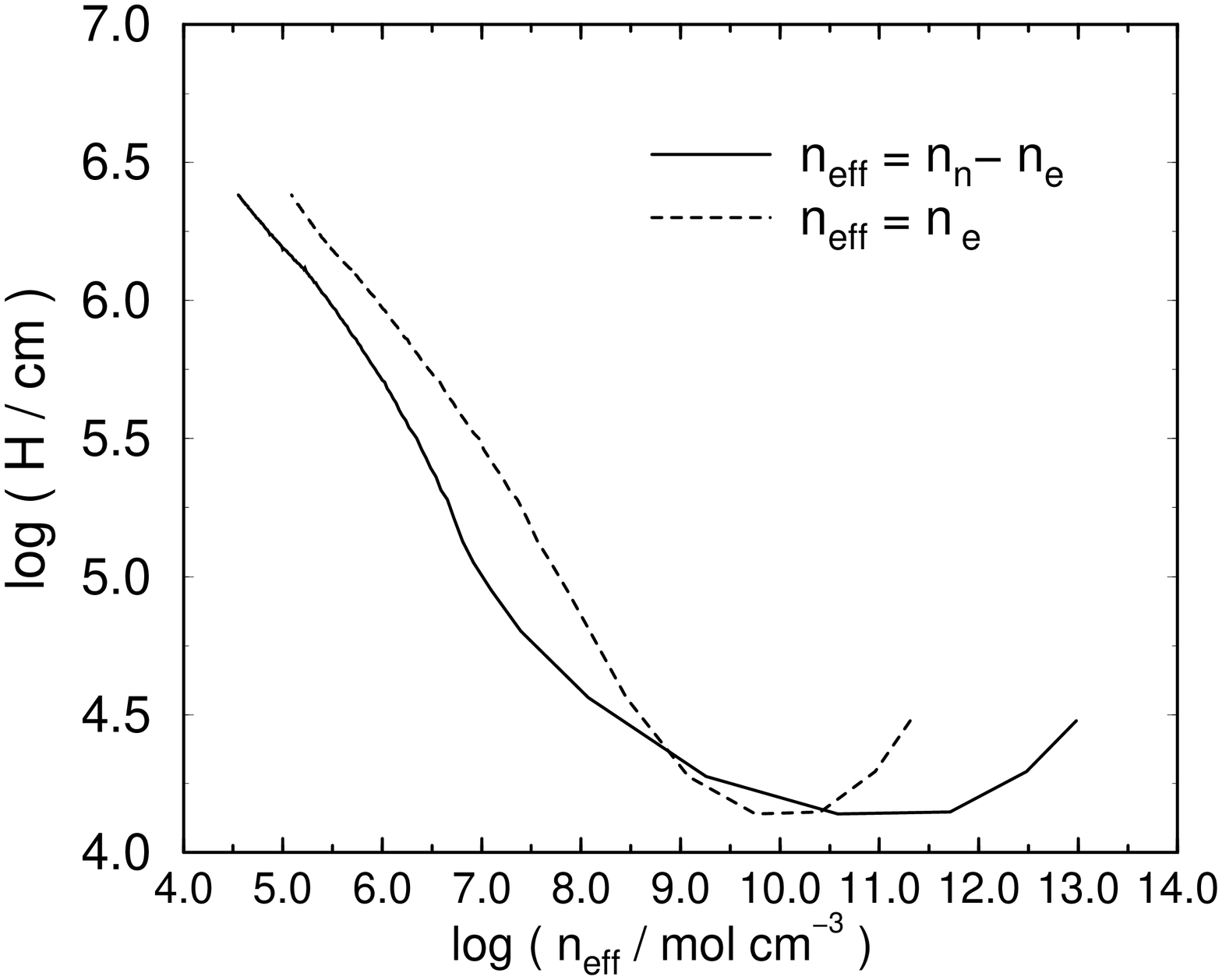,height=16.0cm,width=16.0cm,angle=-90}}
\vglue 2cm
\bf \Huge Fig. 4
\newpage
\centerline{
\psfig{file=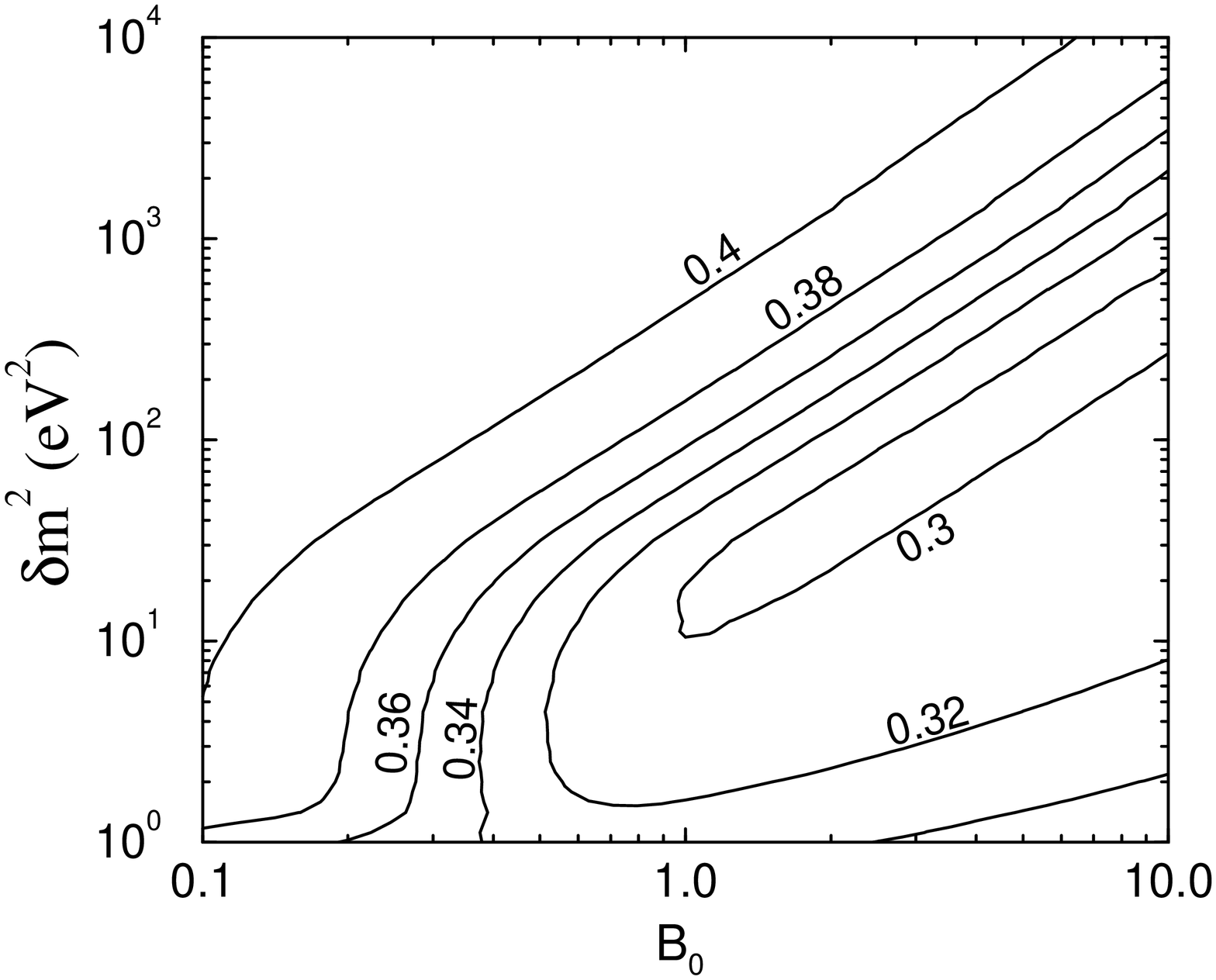,height=16.0cm,width=16.0cm,angle=-90}}
\vglue 2cm
\bf \Huge Fig. 5
\newpage
\centerline{
\psfig{file=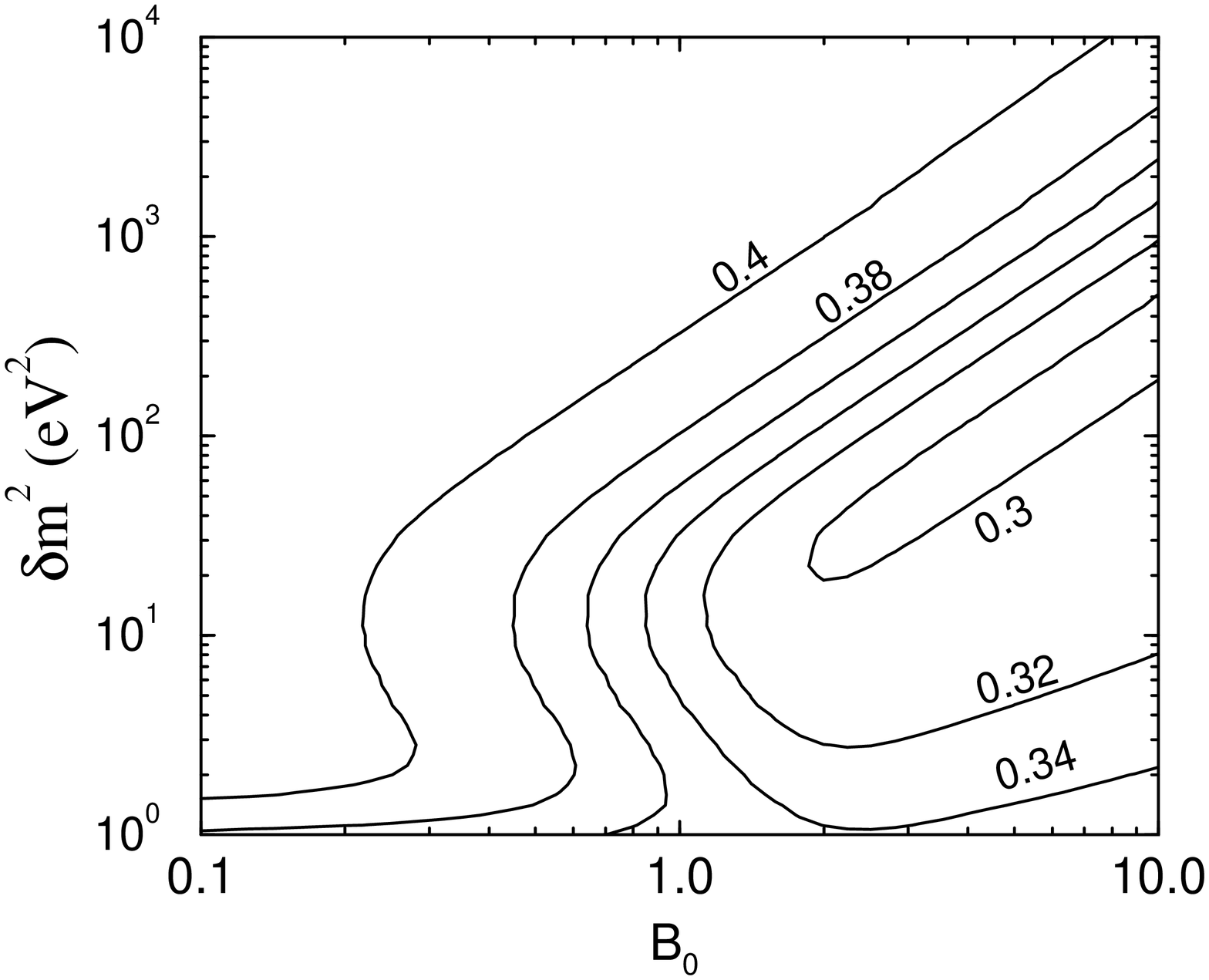,height=16.0cm,width=16.0cm,angle=-90}}
\vglue 2cm
\bf \Huge Fig. 6
\newpage
\centerline{
\psfig{file=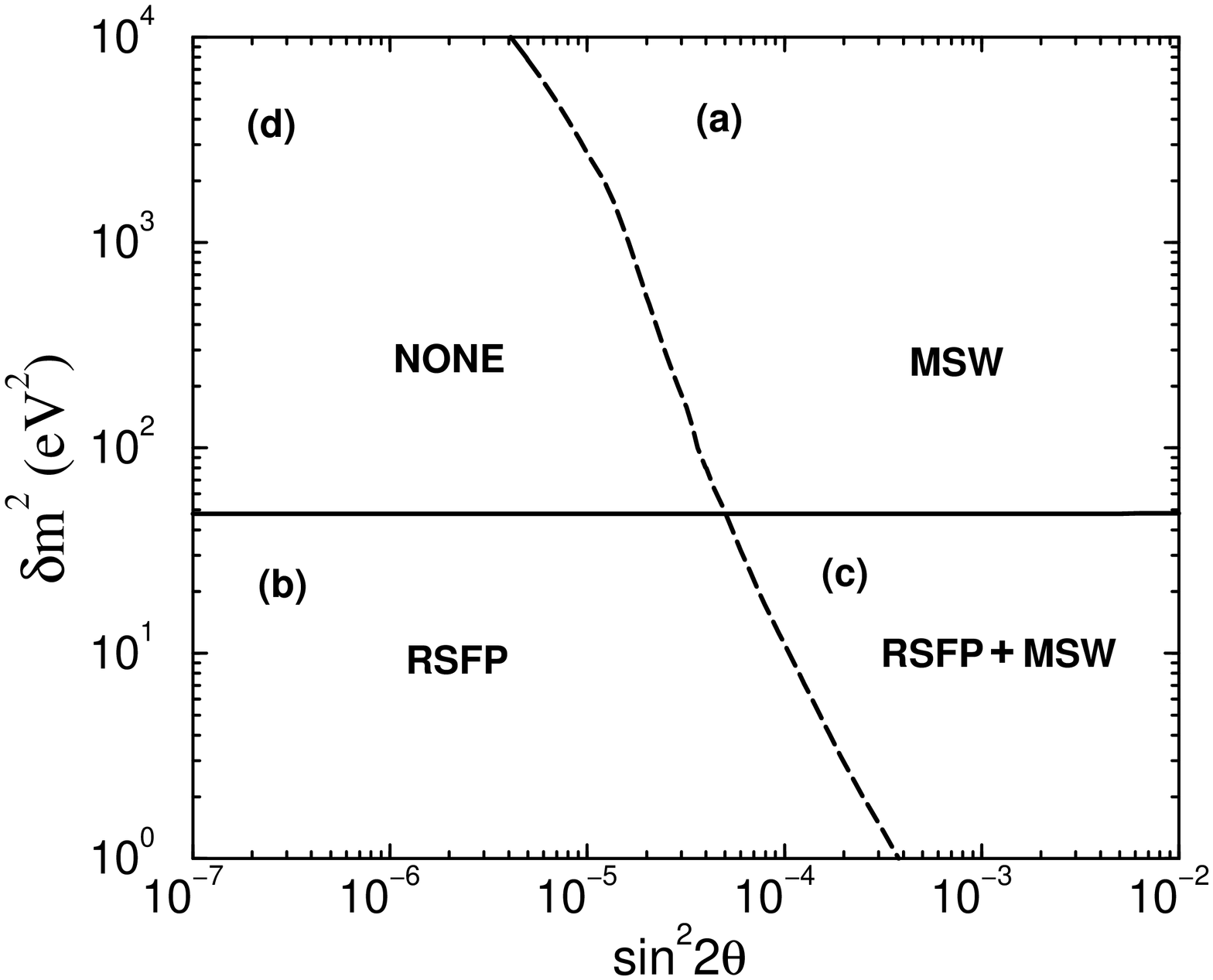,height=16.0cm,width=16.0cm,angle=-90}}
\vglue 2cm
\bf \Huge Fig. 7
\newpage
\centerline{
\psfig{file=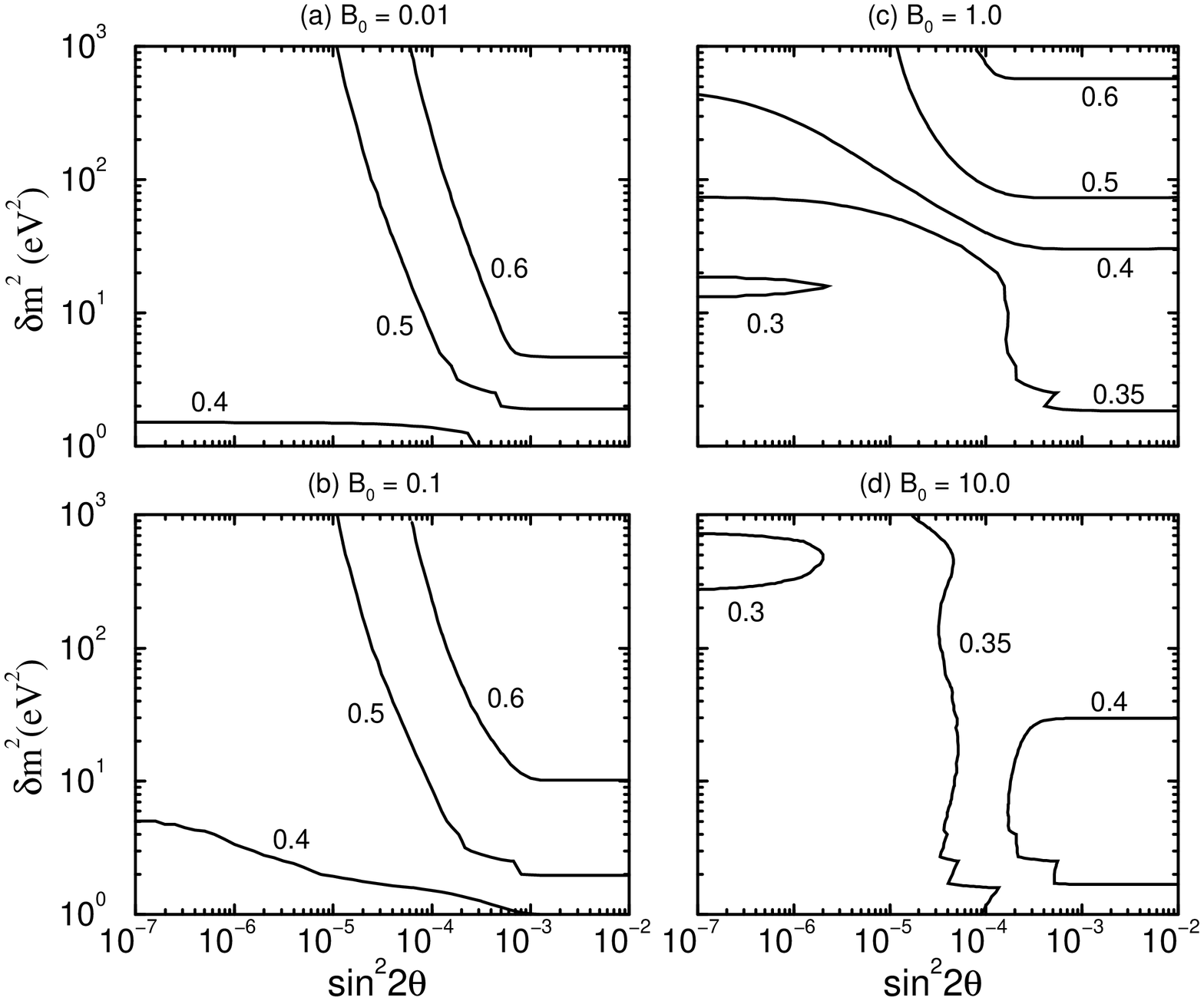,height=19.0cm,width=18.0cm,angle=-90}}
\vglue 2cm
\bf \Huge Fig. 8
\newpage
\centerline{
\psfig{file=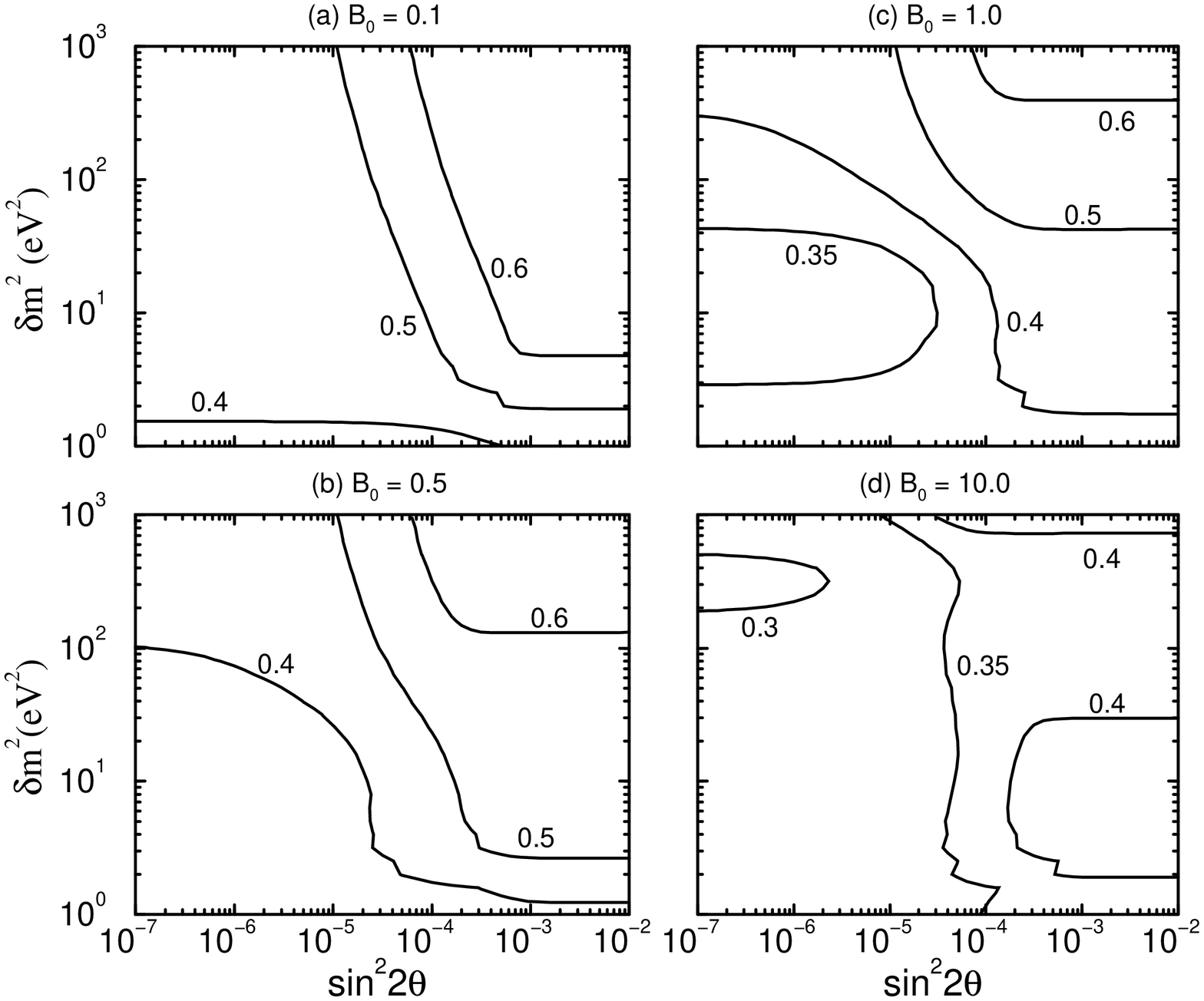,height=19.0cm,width=18.0cm,angle=-90}}
\vglue 2cm
\bf \Huge Fig. 9
\newpage
\centerline{
\psfig{file=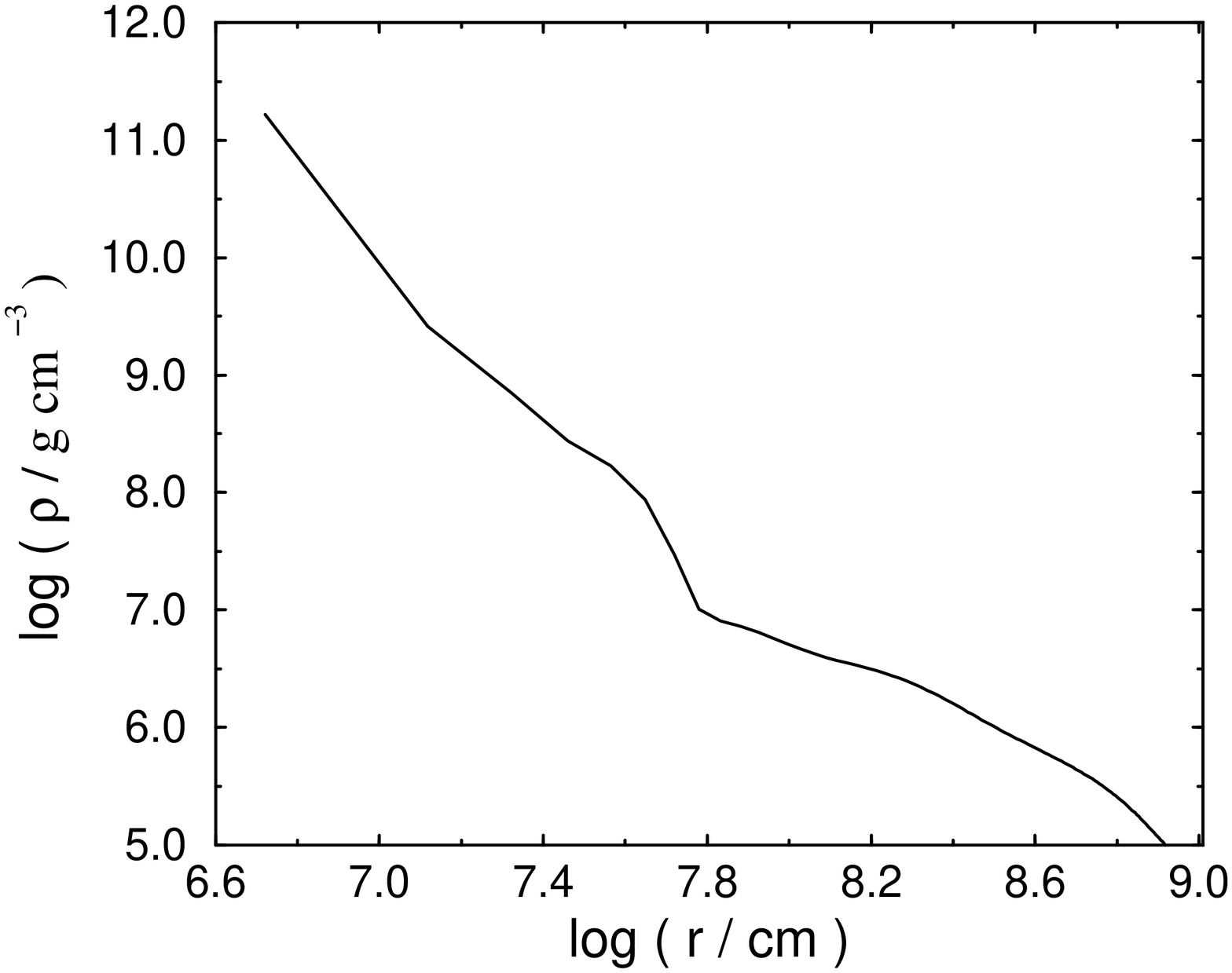,height=16.0cm,width=16.0cm,angle=-90}}
\vglue 2cm
\bf \Huge Fig. 10
\newpage
\centerline{
\psfig{file=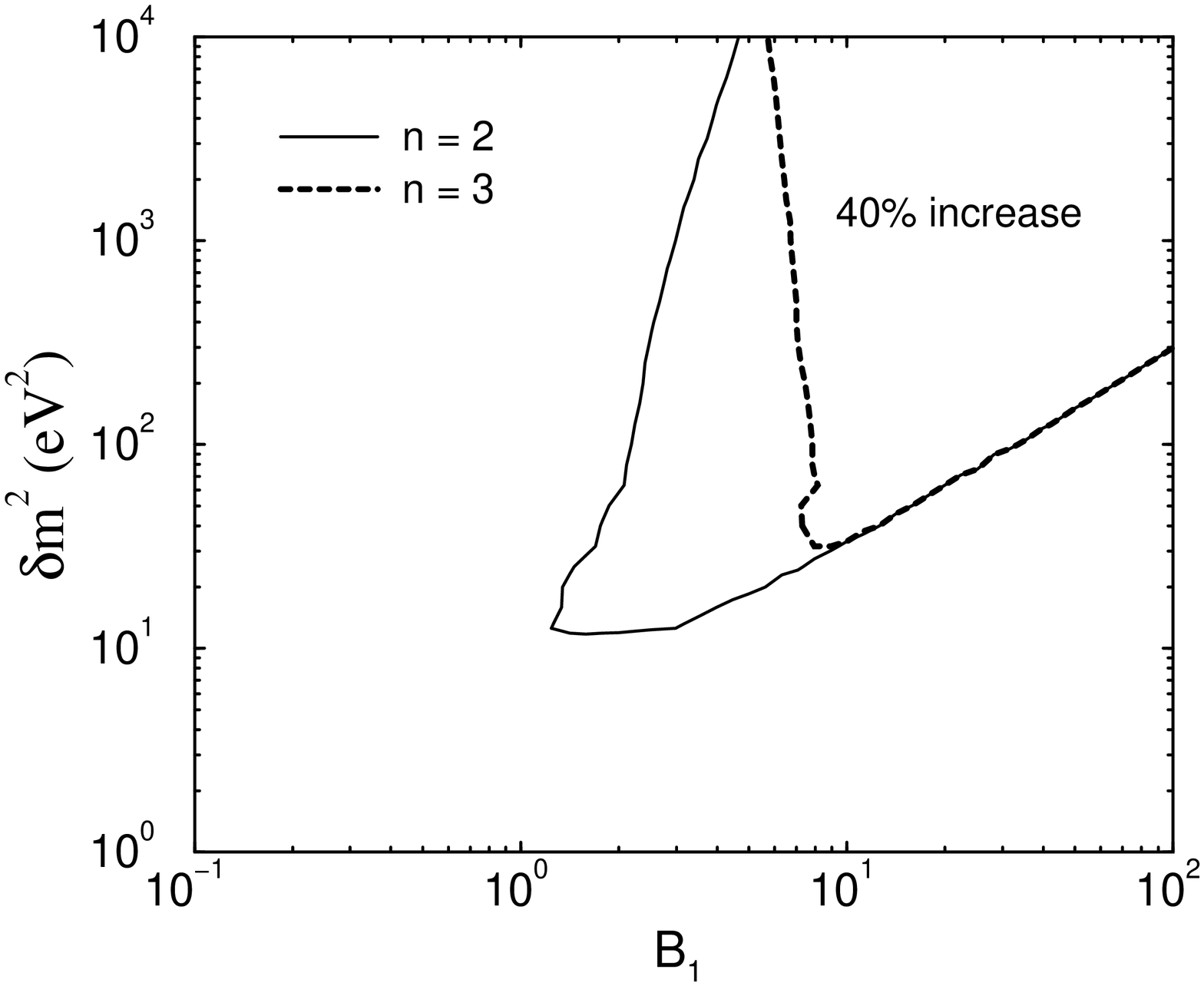,height=16.0cm,width=16.0cm,angle=-90}}
\vglue 2cm
\bf \Huge Fig. 11

\end{document}